\documentclass[iop,twocolappendix]{emulateapj}
\pdfoutput=1
\usepackage{prettyref}
\usepackage{float}
\usepackage{rotfloat}
\usepackage{graphicx}
\usepackage{color}
\usepackage{multirow}
\usepackage{amssymb,amsmath}
\usepackage{sidecap}
\usepackage{bm}
\usepackage{xcolor}
\usepackage{soul}
\usepackage{CJK}

\usepackage{indentfirst}
\usepackage{leftidx}

\graphicspath{.}

\makeatother

\newcommand{\kms}{\rm ~km~s^{-1}}
\newcommand{\ergcps}{\rm ~erg~cm^3~s^{-1}}
\newcommand{\ergccs}{\rm ~erg~cm^{-3}~s^{-1}}
\newcommand{\ergcs}{\rm ~erg~cm^{-2}~s^{-1}}
\newcommand{\Junit}{\rm ~erg~cm^{-2}~s^{-1}~Hz^{-1}}
\newcommand{\dd}{{\rm d}}

\newcommand{\MII}{\rm M_{\,II}}
\newcommand{\MI}{\rm M_{\,I}}
\newcommand{\UI}{$_{\rm ~I}$}
\newcommand{\UII}{$_{\rm ~II}$}
\newcommand{\Lya}{Ly$\alpha$}
\newcommand{\Ha}{H$\alpha$}
\newcommand{\Hb}{H$\beta$}
\newcommand{\Hc}{H$\gamma$}
\newcommand{\HD}{HD~189733b}
\newcommand{\AArm}{\rm~\AA}

\begin{document}
\begin{CJK*}{UTF8}{gkai}

\title{A Model of the H$\alpha$ and N\MakeLowercase{a} Transmission Spectrum of \HD}

\author{Chenliang Huang (黄辰亮)\altaffilmark{1,2}, Phil Arras\altaffilmark{1}, Duncan Christie\altaffilmark{3}, and Zhi-Yun Li\altaffilmark{1}}
\affil{$^1$Department of Astronomy, University of Virginia, Charlottesville, VA 22904, USA; ch4de@virginia.edu, pla7y@virginia.edu}
\affil{$^2$Dept. of Physics and Astronomy, University of Nevada, Las Vegas, NV 89154}
\affil{$^3$Department of Astronomy, University of Florida, Gainesville, FL 32611, USA}


\begin{abstract}

This paper presents a detailed hydrostatic model of the upper atmosphere of HD 189733b, with the goal of constraining its temperature, particle densities, and radiation field over the pressure range $10^{-4}-10\, \mu \rm bar$, where the observed H$\alpha$ transmission spectrum is produced.
The atomic hydrogen level population is computed including both collisional and radiative transition rates.
The Ly$\alpha$ resonant scattering is computed using a Monte-Carlo simulation.  
The model transmission spectra are in broad agreement with the data.
Excitation of the H(2$\ell$) population is mainly by Ly$\alpha$ radiative excitation due to the large Ly$\alpha$ intensity.
The density of H(2$\ell$) is nearly flat over two decades in pressure, and is optically thick to H$\alpha$.
Additional models computed for a range of the stellar Lyman continuum (LyC) flux suggest that the variability in H$\alpha$ transit depth may be due to the variability in the stellar LyC.
Since metal lines provide the dominant cooling of this part of the atmosphere, the atmosphere structure is sensitive to the density of species such as Mg and Na which may themselves be constrained by observations.
Since the H$\alpha$ and Na D lines have comparable absorption depths, we argue that the center of the Na D lines are also formed in the atomic layer where the H$\alpha$ line is formed.

\end{abstract}


\section{INTRODUCTION}

The first detection of an exoplanetary atmosphere was accomplished via measuring the sodium doublet transit signal of HD 209458b \citep{Charbonneau}.
The atmosphere of \HD\ has also been detected by \Lya\ transit \citep{Lecavelier,Bourrier2013}.
The species O$_{\rm\,I}$, Na\UI, and possibly K\UI\ \citep{Pont,Jensen11} were detected by the Hubble Space Telescope (HST) \citep{Ben-Jaffel, Huitson, Pont}.  
An indication of an extended atmosphere was also found in X-ray by Chandra \citep{Poppenhaeger}.
The \Ha, \Hb, and \Hc\ hydrogen lines, and absorption lines from Na\UI\ and possibly Mg\UI\ were detected in \HD's  atmosphere \citep{Jensen11, Jensen, Cauley2015, Cauley2016, Redfield, Wyttenbach, Khala}, which shows the promise of ground based telescopes in studying exoplanet atmospheres.
An H$_2$O feature has been identified at 3.2$\mu$m during the secondary transit \citep{Birkby}.

The atmosphere of HD 209458b has been modeled in order to compare to the observed H \Lya, O, Si $_{\rm III}$, and Na\UI\ lines \citep{Fortney, KoskinenI, KoskinenII, Lavvas}.
For the purpose of studying the \Lya\ emission spectrum, \citet{Menager} calculated the \Lya\ resonant scattering process in the atmosphere of HD 209458b, based on the atmospheric structure model in \citet{KoskinenI}, and \HD\ based on an unpublished model \citep{Koskinen2011}.  A simulation of \HD 's escaping atmosphere has been performed by \citet{Salz}.

As the hot gas in the upper thermosphere is more weakly bound to the planet, conditions there set the boundary condition for the rate of gas escape \citep{Yelle,Garcia,Murray-Clay} .
Among detected species, \Ha\ is a sensitive probe of the planet's upper atmosphere because the excitation and de-excitation processes for H(2$\ell$), the absorber of \Ha, are strongly dependent on the local particle densities, temperature, and radiation field.  In addition, unlike Ly$\alpha$, the interstellar medium is transparent to \Ha\ and this optical line can be observed with large ground based telescopes.  Therefore, the \Ha\ transmission spectrum is a powerful and economical method to probe the structure of the planet's upper atmosphere.

As this work shows, the temperature, and hence scale height, in the region optically thick to \Ha\ is (approximately) set by a balance of photoelectric heating and line cooling by metal species, mainly Mg\UI\ and Na\UI.  If only photoelectric heating and line cooling from hydrogen were included, the atmosphere would be hotter by $\simeq 2000-3000\ \rm K$ \citep{Christie}, giving transit depths far too large in comparison to observations.  Furthermore, several studies \citep{Garcia,KoskinenI,Lavvas} suggested that the transition from atomic to molecule hydrogen occurs at pressures $P \simeq 10~\mu$bar.  These studies included detailed heating and cooling physics in the molecular layer.  But transmission spectra for the Na D doublet and Mg lines may in principle provide further constraints on atmosphere models around this transition altitude.
In addition, the atmospheric temperature of \HD\ derived from the Na doublet transmission spectrum by \citet{Huitson} and \citet{Wyttenbach} is significantly lower than the modeled upper atmosphere temperature found in \citet{Salz} and \citet{Christie}.
A model of the Na transmission spectrum is required to understand these contradictory results.

Both the \HD\ \Ha\ transmission spectrum observed by \citet{Cauley2016} and the Na D transmission spectrum presented by \citet{Wyttenbach} have the spectral resolution to resolve the line core.  The line center transit depths of both observations are about 1-1.5\%, which means the line core absorption features of both species are mostly contributed by the same region in the atmosphere.
Since the temperature of the molecular layer is below 3000 K, and the molecular hydrogen has a large absorption cross section to \Lya\ photon \citep{Black1987}, both the collisional excitation rate and radiative excitation rate are too small to create enough H(2$\ell$) to absorb the \Ha\ in the molecular layer.
These results suggest that the absorption features of both \Ha\ and Na near line center are tracing the atomic layer in the \HD\ atmosphere.

The \HD\ \Ha\ transmission spectrum was modeled by \citet{Christie}, who constructed a hydrostatic atmosphere model similar to the one considered here. In that work, a detailed treatment of \Lya\ radiation transfer was not included, and hence the role of \Lya\ excitation deep in the atmosphere was not appreciated. \citet{Christie} showed that if collisional excitation dominates, it would lead to a fairly constant H(2s) density within the atomic layer, because of the combination of increasing temperature and decreasing H(1s) density with radius.  
In attempting to improve on their model, it was found that \Lya, especially from recombinations occurring within the atmosphere, could give radiative excitation rate to H(2p) much larger than the collisional excitation rate, and that this excitation could occur deeper in the atmosphere where the H(1s) density is higher, even though the temperature is much lower there. This key insight motivated the detailed \Lya\ radiative transfer treatment in the present work.

\section{ analytic estimates }
\label{sec:analytic}
\begin{table}
  \centering
  \caption{Adopted values for the orbital and physical parameters of HD~189733 and \HD}
  \begin{tabular}{cc}
    \hline  \hline
    Stellar type & K2V \\ 
    Star mass  & $M_\star = 1.60\times 10^{33}\, \rm g$  \\
    Star radius & $R_\star = 5.60\times 10^{10}\, \rm cm$  \\
    Semi-major axis & $a = 0.031\, \rm AU$ \\
    Planet mass & $M_p = 2.17\times 10^{30}\, \rm g$  \\
    Planet radius & $R_p = 8.137 \times 10^9\, \rm cm$  \\
    Planet surface gravity & $g_p = GM_p/R_p^2 = 2.2\times 10^3\, \rm cm\ s^{-2}$ \\
    \hline
  \end{tabular}
  \tablecomments{ Source: exoplanets.org}
  \label{tab:param}
\end{table}

Observations of the \Ha\, transmission spectrum for \HD\ show a $\sim 1 \%$ line center transit depth, and a half-width of $\sim 0.4 \AArm$ (e.g. \citet{Cauley2015}).  \citet{Wyttenbach} measured a nearly $\sim 1\%$ transit depth for both Na D lines, among which the Na\UI\ 5890~\AA\ is slightly deeper.  This section contains analytic estimates for the conditions in the atmosphere required to generate the observed \Ha\ and Na D absorption lines. 

The outermost reaches of HD 189733b's atmosphere are highly ionized by stellar photons. Moving inward, the radiative recombination rate eventually increases to the point that the atmosphere is dominated by atomic hydrogen, at a pressure level $P \simeq 10^{-3}\, {\rm \mu bar}$. The temperature near the transition from ionized to atomic is regulated to be near $T\simeq 10,000\, \rm K$, which is far too hot for molecular hydrogen to form, and hence there must be a layer of atomic hydrogen extending over the temperature range $T \simeq 2,500-10,000\, \rm K$. In the hydrostatic model presented in this paper, the temperature in the atomic layer is set by a balance of photoelectric heating and atomic line cooling, for which H$_2$ will dominate at $P \ga 10\, \mu\, \rm bar$. In terms of size, the atomic hydrogen layer extends over $\sim 10$ pressure scale heights, has a mean molecular weight $\mu \simeq 1.3$ and mean temperature $T \simeq 5,000\, \rm K$. As compared to the underlying molecular layer, with mean molecular weight $\mu \simeq 2.3$ and temperature $T \simeq 1,000\, {\rm K}$, the scale height in the atomic layer is larger by a factor of $\sim 10$ compared to the molecular layer, and hence can give rise to absorption to much larger altitudes.

The origin of the H(2$\ell$) population requires a detailed level population calculation. In the present model of the atomic layer, it is found that radiative excitation by \Lya\, creates a nearly constant H(2$\ell$) density over $\sim 6$ pressure scale heights near the base of the atomic layer. This is the cause of the \Ha\, absorption. 
The underlying molecular layer is expected to be optically thin to \Ha\, for two reasons.
First, the density of atomic hydrogen drops rapidly into the molecular layer (e.g. \citealt{Lavvas}), as compared to the base of the atomic layer, due to the much lower temperature there. 
Second, the mean free path to true absorption of \Lya\, by H$_2$ \citep{Black1987} rapidly decreases as the H$_2$ density increases, and hence the \Lya\, intensity is expected to drop rapidly in the molecular layer, with an associated decrease in excitation to the n=2 state. 

The measured line center transit depth $\Delta F/F \sim 1 \%$ requires a certain area to be optically thick. The scale height in the atomic layer is
\begin{eqnarray}
H & = & \frac{k_B T}{\mu m_p g} =   1500\, {\rm km} \left( \frac{T}{5000\, \rm K} \right),
\end{eqnarray}
where the mean molecular weight has been assumed to be $\mu=1.3$.
If N scale heights are optically thick, as compared to the neighboring continuum radiation, this gives an extra absorption depth
\begin{eqnarray}
\frac{\Delta F}{F} & \simeq & N \frac{2\pi R_p H}{\pi R_\star^2}  =  8 \times 10^{-4} N \left( \frac{T}{5000\, \rm K} \right) 
\nonumber \\ & = & 0.5\% \ \left( \frac{N}{6} \right)
\left( \frac{T}{5000\, \rm K} \right).
\end{eqnarray}
Hence the measured line center depth can only be explained by a layer extending many pressure scale heights, and with high temperature $T \ga 5,000\, \rm K$.

The line center optical depth must be greater than unity over the above annulus.
As will be shown, $n_{2\ell}$ is nearly constant over a large pressure range.
Then for an effective path length $2\sqrt{2 N R_p H} \simeq 7\times 10^9\ {\rm cm}\ (N/6)^{1/2}(T/5,000\rm K)^{1/2}$
and \Ha\, line center cross section $\sigma_0=5 \times 10^{-13}\ {\rm cm^2}\ (5,000\rm K/T)^{1/2}$, the maximum line center optical depth is
\begin{eqnarray}
\tau_0 & \simeq &  35 \times \left( \frac{ n_{2\ell} }{ 10^4\ \rm cm^{-3} } \right)
\end{eqnarray}
for the fiducial value $n_{2\ell} = 10^4\ \rm cm^{-3} $.

The line width is mainly set by the temperature and the maximum line center optical depth. For sightlines optically thick to \Ha\, at a line center, the optical depth at distance $x=(\nu-\nu_0)/\Delta \nu_D$ from line center is $\tau(x) = \tau_0 \exp(-x^2)$, when Doppler broadening dominates. All frequencies out to $x \simeq \sqrt{ \ln (\tau_0) }$ are then optically thick. In velocity units, the
line width is then 
\begin{eqnarray}
\Delta v & = & \left( \frac{2k_BT}{m_p} \right)^{1/2} \left( \ln \tau_0 \right)^{1/2} = 9.1\, {\rm km\ s^{-1}}\left( \ln \tau_0 \right)^{1/2}
\end{eqnarray}
or in wavelength units
\begin{eqnarray}
\Delta \lambda & = & 0.2\AArm\, \left( \ln \tau_0 \right)^{1/2}.
\end{eqnarray}
Since $\tau_0 \gg 1$ for \Ha\, over a large region, the width will be larger than the thermal width.

If \Lya\ excitation is balanced by radiative de-excitation, and $\ell$-mixing populates the 2s state,  the abundances relative to the ground state are
\begin{eqnarray}
\frac{ n_{2p} }{ n_{1s} } & \simeq & 3 \frac{ n_{2s} }{ n_{1s} }  =  \frac{ J_{12} B_{12}}{A_{21}} 
\nonumber \\ & \simeq & 
 10^{-8}\, \left( \frac{J_{12}}{10^{-9}\Junit} \right),
\end{eqnarray}
where $A_{21}$ and $B_{12}$ are the Einstein A and B (absorption) coefficient respectively, following the definition in \citet{Rybicki}.
The peak \Lya\, intensity $J_{12} \simeq 0.1 F_{\rm LyC} / \Delta \nu_D \simeq 10^{-8} \Junit$ is found near the peak in photoionization of ${\rm H}(1s)$ near $P \simeq 10^{-3}\, \rm \mu bar$. Here, $F_{\rm LyC} \simeq 10^4\, \ergcs$ is the LyC flux deposited in that region, and it is assumed that each ionization is balanced by a recombination producing a \Lya\, photon. While the ratio of excited state to ground state is high near the peak in \Lya, the H(1s) density there is too small for significant H(2$\ell$) density. 

The key question for the H(2$\ell$) population is how fast the \Lya\, intensity decreases moving deeper into the atmosphere. If \Lya\, intensity does not drop off too fast, the rapid increase in H(1s) density with depth will lead to higher H(2$\ell$) density deeper in the atmosphere. One can imagine two limits to answer this question.  In the first limit, there is a shallow source of \Lya\, at optical depth $\tau_s$ and the intensity $J_\nu(x,\tau)$ is desired at $\tau \gg \tau_s$. An analytic solution based on the Fokker-Planck equation given in \citet{Harrington} is
\begin{eqnarray}
J_\nu(x,\tau) & \simeq & 0.1 \left( \frac{ F_0 }{ \Delta \nu_D} \right) \left( \frac{\tau_s}{\tau} \right).
\end{eqnarray}
This expression is valid in the plateau of the intensity near line center.
Since \Lya\, optical depth $\tau \propto n_{1s}$, this scaling for $J_\nu$ would give $n_{2\ell} \propto n_{1s} J_\nu \simeq \rm constant$ with depth.
The second limit to imagine is where radiation is emitted and absorbed locally, which is appropriate deep in the atmosphere where $\tau \sim 10^8$ scatterings are required to escape the atmosphere.
For a constant, frequency-integrated source function $S$, and true absorption by metal species, the frequency-integrated photon energy equation becomes $S \simeq n_m \sigma_m J$, where $J$ is the frequency-integrated intensity, $n_m$ is the metal number density and $\sigma_m$ is the metal photoionization cross section at \Lya. For a constant mixing ratio, $n_m \propto n_{1s}$,
and again $n_{2\ell} \simeq \rm constant$. While the scaling found by these two estimates, constant 2$\ell$ density, is the same, it is found that the local balance of sources and sinks is the applicable limit in the present atmosphere model.

Up to this point, the estimates have been concerned with the \Ha\ transmission line, however, the center of each line in the Na doublet may also be formed in the atomic layer.
Because the cross section of Na\UI\ 5890 is larger than Na\UI\ 5896 by a factor of 2, the difference in transit radius between the two resolved line centers corresponds to $\ln (2) H$, assuming a constant Na\UI\ number fraction.
Keeping in mind the error bars in the measurement, according to \citet{Wyttenbach}, the difference in transit radius between the Na D lines is $\sim 3000$ km, which gives a local scale height $H\simeq 4300 ~\rm km$.  Plugging in the transit radius $R({\rm Na})=9.4 \times 10^9~\rm cm$ to compute the gravity, assuming $\mu=1.3$, the temperature derived from the line centers is $T\simeq 11,000~\rm K$.  In order to explain this scale height with a molecular gas, the temperature has to be higher by a factor of 2.  But this high temperature is inconsistent with the gas being in molecular form.

\section{hydrogen level population}
\label{sec:level}

Balmer line photons are absorbed by the 2$\ell$ excited states of H. Due to ionization and the subsequent recombination cascades, and a radiation excitation temperature different from the gas temperature, the level populations are not set by the Boltzmann distribution at the local gas temperature. 
Therefore, a study of the H level population over the range of densities, temperatures, and intensities found in hot Jupiter atmosphere is required. The following processes are considered.

\begin{enumerate}
\item Hydrogen radiative (de-)excitation of all possible electric dipole transitions between multiplets up to $n=6$ ~\citep{Wiese}; 

\item Electron collisional (de-)excitation for transitions from $1s$ to each sub-state $\ell$ up to $n=5$, from $2s$ to each sub-state $\ell$ up to n=5, and from $2p$ to each sub-state $\ell$ up to n=3 (CHIANTI database, \citet{Chianti,Chianti8}); 

\item Electron collisional $\ell$-mixing between $2s$ and $2p$~\citep{Seaton55};

\item Proton collisional $\ell$-mixing between $2s$ and $2p$~\citep{Seaton55}, and $\ell$-mixing for levels $3\leq n\leq 6$~\citep{Vrinceanu};

\item Electron collisional ionization and three-body recombination for each sub-state $\ell$ up to n=4 and total ($\ell$-unresolved) rates for $n=$5 and 6~\citep{Janev}.  The cross sections are assumed to be equal for all sub-states in the same level for $n=$5 and 6;

\item Photoionization of each sub-state $\ell$ up to n=6~\citep{TOP}.  The corresponding recombination rate can be calculated using the Milne relation.
\end{enumerate}

\begin{figure}
\begin{center}
\includegraphics[width=0.5\textwidth]{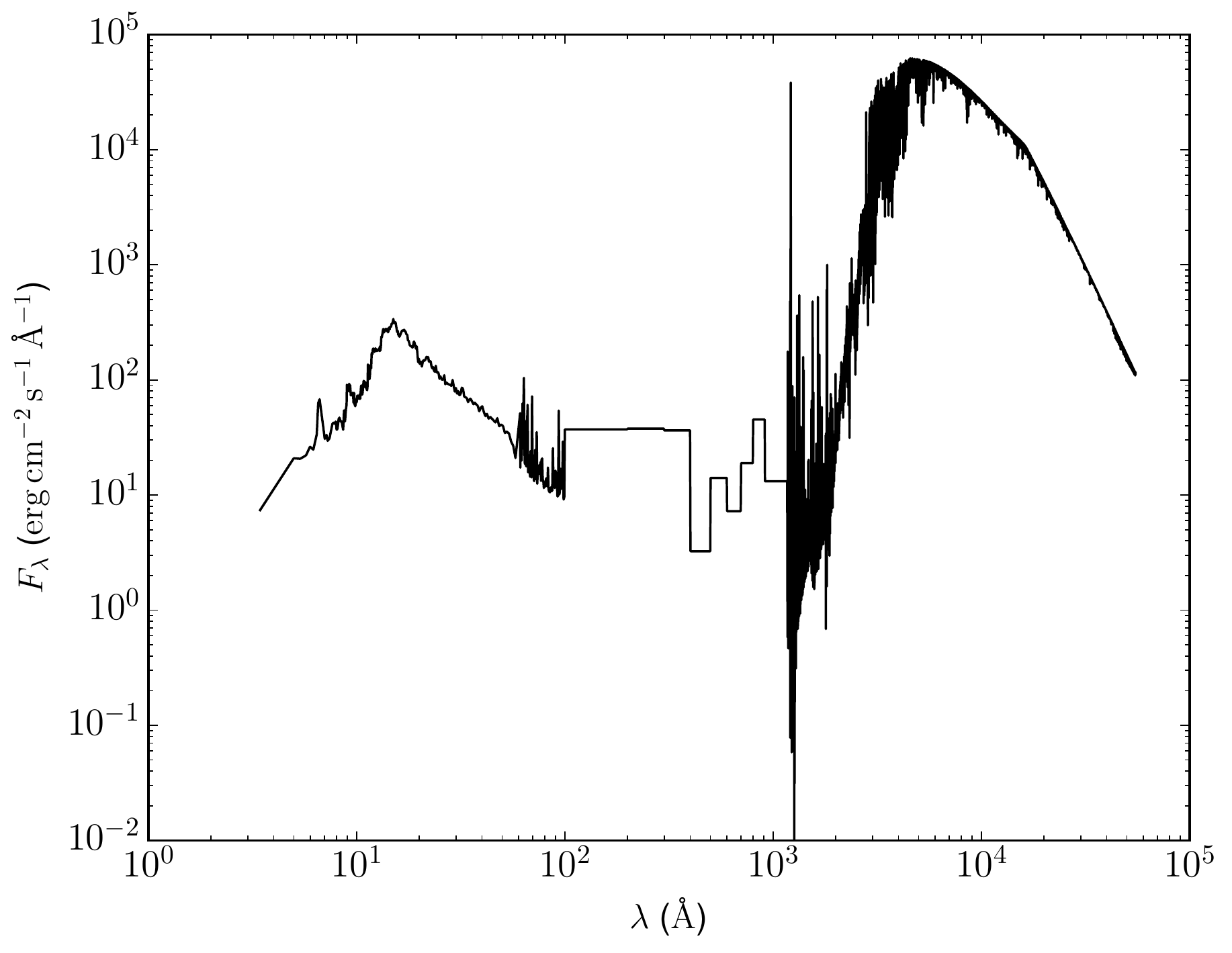}
\caption{Stellar energy flux spectrum ($F_\lambda$) against wavelength ($\lambda$) used in the model at orbital distance of HD 189733b.}
\label{spec}
\end{center}
\end{figure}

The stellar spectrum of $\epsilon$ Eri, another K2V star, given by the MUSCLES treasury survey version 2.1 \citep{France,Yongblood,Loyd} is applied in the model.  Its energy flux ($F_\lambda$) is normalized to the distance $a=0.031\ \rm AU$ of HD 189733b from its parent star, shown in Figure~\ref{spec}. 
The \Lya\ fluxes are combined with models of solar active regions \citep{Fontenla} to estimate the EUV luminosity in the wavelength region 100 -- 1170\AArm\ in 100\AArm\ bins.
The stellar flux attenuation due to bound-bound transitions is not included in this calculation because it is relatively unimportant in setting the level populations, and performing radiation transfer calculations in addition to \Lya\ would greatly complicate the analysis.
Bound-bound transitions due to Ly$\beta$, Ly$\gamma$ etc. are ignored as it is assumed that these photons rapidly down-convert into a \Lya\ photon and lower series photons at the large optical depths of interest for these lines.

The level population is determined by the kinetic equilibrium between production and loss processes. The equation of rate equilibrium for the state $j$ is
\begin{multline}
 \alpha_{j} n_e n_p + \beta_{j}n_e^2n_p + \sum_{k}C_{k\to j}^{(e)}n_en_{k} \\
+ \sum_{k}C_{k\to j}^{(p)}n_pn_k + \sum_{k>j}A_{k\to j}n_k + \sum_{k<j}B_{k \to j}\bar{J}_{k \to j}n_k \\
= \Big(\Gamma_{j} + \sum_{k}C_{j \to k}^{(e)}n_e + \sum_{k}C_{j \to k}^{(p)}n_p\\
 + C_{j\to \infty}^{(e)}n_e + \sum_{k>j}B_{j \to k}\bar{J}_{j \to k} + \sum_{k<j}A_{j\to k} \Big) n_{j},
 \label{eq:levelpopeqn}
\end{multline}
where $\alpha_{j}$ and $\beta_{j}$ denote radiative and three-body recombination rate coefficients for state $j$, respectively.   Case B recombination with $\alpha_{1s}\equiv 0$ is employed. The $n_k$ is the number density of hydrogen in sub-state $k$.
The inequality $k>j$ denotes a downward transition from a state with principle quantum number of state $k$ larger than that of state $j$. The rate coefficients
$C_{j\to k}^{(e/p)}$ are for electron/proton impacts causing a transition from state $j$ to state $k$, and the state $\infty$ represents ionization. The rate for proton collisions is only included for the $\ell$-mixing transitions at fixed principal quantum number.  The spontaneous radiative decay rates are $A_{j\to k}$.
The photoionization rate from state $j$ is denoted $\Gamma_{j}$, and attenuation from the overlying gas is included in the ground state photoionization calculation. The optical depth for ionizing photons is computed as
\begin{equation}
 \tau_{{\rm att},\nu}=\sigma_{1s}(\nu)N_{1s}+\sum_{m} \sigma_{m}(\nu) N_{m},
\end{equation}
where $N_{k}(r)=\int_r^\infty \dd r' n_{k}(r')$ is the column of species $k$ above the layer under consideration. The sum over the subscript $m$ stands for all metal species considered. As discussed in Section \ref{hydro}, this is $m=$neutral and first ionized C, O, Mg, Si, and S, and neutral Na and K.
Including the attenuation factor $e^{-\tau_{\rm att},\nu}$, the photoionization rate is
\begin{equation}
  \label{eq:12}
\Gamma_{j,\rm pi}=\int_{\nu_{j,\rm th}}^\infty \sigma_{j,\rm pi}(\nu)\frac{F_{\nu}}{h\nu}e^{-\tau_{{\rm att},\nu}}\dd \nu,
\end{equation}
where $\nu_{\rm th}$ is the corresponding photoionization threshold frequency.  The photoionization rates and photoelectric heating rates (see Equation~\ref{eq:7} and~\ref{eq:8}) at $\tau_{\rm att}=0$ are listed in Table~\ref{tab:pi_rates}.

The H(1s) state can also undergo ``secondary ionization" by photoelectrons generated  when a photon with much higher energy than the ionization threshold ionizes a hydrogen or metal atom.  If $n_e/n_{1s}$ is small, photoelectrons can cause H(1s) collisional ionization and excitation before sharing their energy with other electrons through Coulomb collisions, which would increase the photoionization rate and reduce the photoelectron heating efficiency.
Different from the treatment of constant efficiency applied in \citet{Yelle} and \citet{Murray-Clay}, or the ionization fraction $x_e=n_e/n_{1s}$ independent efficiency applied in \citet{KoskinenI}, an efficiency dependent on local $x_e$ and photoelectron energy $E=h(\nu-\nu_{\rm th})$ is used here.  Given the incoming photon frequency $\nu$ and ionization threshold energy $E_{\rm th}$, \citet{Draine} gives the number of secondary ionizations per ionizing photon, in the case of $E>50 \text{ eV} \text{ and } x_e<1.2$, to be
\begin{equation}\label{ke}
k_e(E)=\left(\frac{E-15{\rm\ eV}}{35{\rm\ eV}}\right)\left( \frac{1-x_e/1.2}{1+18x_e^{0.8}/\ln(E/35{~\rm eV})}\right).
\end{equation}
After correcting for the secondary ionizations, the form for the H(1s) ionization rate which is used in the rate equations becomes
\begin{align}
  \Gamma_{1s} & =\Gamma_{1s}^{\rm 1st }+\Gamma_{1s}^{\rm 2nd} \nonumber \\
  & =\int_{\nu_{\rm th}}^\infty \sigma_{1s}(\nu)(1+k_e)\frac{F_\nu}{h\nu}e^{-\tau_{{\rm att},\nu}}\dd \nu,
  \label{eq:Gamma1s}
\end{align}
where $\Gamma_{1s}^{\rm 1st }$ and $\Gamma_{1s}^{\rm 2nd}$ stand for primary and secondary photoionization rate from H(1s) respectively.
Secondary photoionization becomes an important consideration at pressures $P \ga 0.1\, \mu$bar, where the initially more abundant lower energy photons have already been absorbed higher in the atmosphere, and the dominant photons being absorbed can induce at least one secondary ionization on average.

\begin{table}
  \caption{Photoionization rates and photoelectric heating rates}
  \begin{center}
    \begin{tabular}{ccc}
      \hline  \hline
      Species & $\Gamma_{\rm pi} $ (s$^{-1}$) & $Q_{\rm pi}$ (erg s$^{-1}$)  \\
      \hline
      H($1s$) & $1.61\times 10^{-3}$ & $5.91\times 10^{-15}$ \\
      H($2s$) & 25.7 & $1.42\times 10^{-11}$ \\
      H($2p$) & 21.5 & $1.13\times 10^{-11}$ \\
      O\UI & $3.06\times 10^{-3}$ & $6.14\times 10^{-14}$ \\
      O\UII & $7.10\times 10^{-4}$ & $1.51\times 10^{-14}$ \\
      C\UI & $7.43\times 10^{-3}$ & $5.91\times 10^{-14}$ \\
      C\UII & $3.90\times 10^{-4}$ & $1.07\times 10^{-14}$ \\
      Mg\UI & $6.12\times 10^{-4}$ & $2.96\times 10^{-14}$ \\
      Mg\UII & $2.23\times 10^{-4}$ & $2.14\times 10^{-14}$ \\
      Si\UI & $2.18\times 10^{-2}$ & $1.02\times 10^{-14}$ \\
      Si\UII & $2.07\times 10^{-4}$ & $1.07\times 10^{-14}$ \\
      S\UI & $2.21\times 10^{-2}$ & $1.56\times 10^{-13}$ \\
      S\UII & $2.30\times 10^{-4}$ & $1.04\times 10^{-14}$ \\
      Na\UI & $9.92\times 10^{-4}$ & $5.36\times 10^{-14}$ \\
      K\UI & $2.08\times 10^{-3}$ & $2.60\times 10^{-14}$ \\
      \hline
    \end{tabular}
  \end{center}  
  \tablecomments{ Atmospheric attenuation, secondary ionization effect and the contribution by \Lya\ photon are not included.}
  \label{tab:pi_rates}
\end{table}

Lastly, the bound-bound radiative excitation rates are given by
\begin{equation}
B_{l \to u}\bar{J}_{l \to u}=\frac{g_u}{g_l} \frac{c^2 A_{u\rightarrow l}}{2 h \nu^3} \int \dd \nu J_\nu \phi(\nu),
\end{equation}
where $\bar{J}_{l \to u}$ is line profile weighted mean intensity and $\phi(\nu)$ is the Voigt profile. 

\begin{table}
  \caption{Rate which may be important to $n_{2p}$ population and de-population}
  \centering
  \begin{tabular}{*2c}
    \hline  \hline
    Process & Rates ($\rm cm^{-3}s^{-1}$) \\
    \hline
    Radiative excitation $1s \to 2p$ & $1.7 \times 10^{11} (n_{1s}/10^{10}~\rm cm^{-3})$ \\
    Collisional excitation $1s \to 2p$ & $8.4\times 10^{4} (n_e n_{1s} /10^{19}~\rm cm^{-6})$ \\
    Radiative recombination to $2p$ &  $6.2\times 10^{4} (n_e n_{p} /10^{18}~\rm cm^{-6})$ \\
    Spontaneous decay $2p \to 1s$ &  $6.3\times 10^{11} (n_{2p} /10^{3}~\rm cm^{-3})$ \\
    $p$ collisional $\ell$-mixing $2p \to 2s$ &  $1.9 \times 10^{8} (n_{2p}n_p/10^{12}~\rm cm^{-6})$ \\
    Photoionization from $2p$ &  $2.2\times 10^{4} (n_{2p} /10^{3}~\rm cm^{-3})$ \\
    Collisional de-excitation $2p \to 1s$ &  $7.4\times 10^{3} (n_{2p} n_e/10^{12}~\rm cm^{-6})$ \\
    Collisional ionization from $2p$ &  $7.4\times 10^{2} (n_{2p} n_e/10^{12}~\rm cm^{-6})$ \\
    \hline
  \end{tabular}
  \tablecomments{In the table, $\bar{J}_{Ly\alpha}=2\times 10^{-9}\Junit$ and $T$=8000 K are applied.  The reference numbers are $n_{1s}=10^{10}~\rm cm^{-3}$, $n_e=n_p=10^{9}~\rm cm^{-3}$ and $n_{2p}=10^{3}~\rm cm^{-3}$.}
  \label{tab:n2p_rates}
\end{table}

Equation \ref{eq:levelpopeqn} is evaluated for all $0 \leq \ell \leq n-1$ and $1 \leq n \leq 6$, resulting in a linear system of 21 equations in total for the number density of each $(n,\ell)$ state, $n_{n\ell}$. The quantities $n_e$, $n_p$, and $T$ are treated as given parameters in the equations.  The linear system is solving using Gauss-Jordan elimination~\citep{Recipes}.

The rates of important processes related to the $2p$ state population are listed in Table~\ref{tab:n2p_rates}.  The rates related to higher excited states are not listed in the table because they cannot have a large net effect on $2p$ in steady state, unless the higher state itself has a large source or sink, which is not the case.  The stellar \Lya\ mean intensity (see Section~\ref{MC}) is applied for the estimate.  The table shows that all other rates except the $\ell$-mixing rates between $2s$ and $2p$ are negligible compared to the radiative rates between $2p$ and $1s$.  The proton collisional $\ell$-mixing rate is much larger than the electron rate.  The $\ell$-mixing rates between $2s$ and $2p$ nearly cancel each other and $2\ell$ states are in collisional equilibrium due to the large $\ell$-mixing rates at the densities of interest.  Thus, $A_{2p\to 1s}n_{2p}$ and $B_{1s\to 2p}\bar{J}_{Ly\alpha}n_{1s}$ completely dominate the $2p$ generation and destruction rate, and the n=2 state number densities are simply set by $\bar{J}_{Ly\alpha}$, so we have,
\begin{equation}
  \label{eq:6}
  n_{2p}\approx 3n_{2s}\approx n_{1s} \frac{g_u}{g_l}\frac{c^2\bar{J}_{Ly\alpha}}{2h\nu^3}.
\end{equation}
In Section~\ref{sec:fiducial}, it will be shown that this is a good approximation for the whole simulation region.  To obtain the intensity of Ly$\alpha$, a resonant scattering study of \Lya\ photon will be discussed in Section~\ref{MC}.

\section{the atmosphere model}\label{hydro}

\subsection{Basic Structure}

Following \citet{Christie}, a spherically-symmetric, hydrostatic atmosphere model is constructed for the region composed of ionized and atomic gas sitting above the molecular atmosphere. 
The transit radius measured in broadband optical wavelengths is $R_p$, and the base of the atomic layer is at radius $R_b > R_p$. The thickness of the molecular layer below the atomic layer is then $R_b-R_p$.  Assuming an isothermal molecular layer with equilibrium temperature $T=T_{\rm eq}=1140{\rm~K}$ \citep{Wyttenbach}, the thickness is 
\begin{equation}
R_b-R_p\approx \frac{k_{\rm B} T_{\rm eq} R_p^2}{\mu m_{\rm H}GM_p}\ln \left( \frac{P_p}{P_b}\right),
\end{equation}
where $\mu \approx 2.3$ is the mean molecular weight, $P_p=1\, $bar is the pressure of the optical photosphere suggested by~\citet{Sharp}.  It is assumed that the pressure at the base of the atomic layer is $P_b=10\, \mu$bar. It will be shown that the temperature in the atomic layer becomes small enough for molecular hydrogen to dominate there, for the assumptions used here. 

\subsection{Differential Equations}

Given the temperature and number density of each species at one level in the atmosphere, the equation of hydrostatic balance and equations for the column of each species must be integrated inward to find pressure and columns at the next step inward.
The hydrostatic balance equation is
\begin{align}
\frac{\dd P}{\dd r} & =-\frac{\rho GM_p}{r^2}
\label{eq:hb}
\end{align}
and the columns are integrated as
\begin{align}
\frac{\dd N_i}{\dd r} & = -n_i.
\label{eq:Ni}
\end{align}
The subscript $i$ stands for each species, including H(1s) and the individual neutral and first ionized metal element considered.  
The ideal gas law for the gas pressure is
\begin{equation}
P=(n_e+(n_p+n_{\rm H})(1+f_z+f_{\rm He}))k_{\rm B}T,
\label{eq:P}
\end{equation}
where $f_z=\sum f_m$ is the sum of metal species relative number density abundance to hydrogen, and $f_{\rm He}$ is the fraction of He by number assuming solar abundance \citep{Asplund}.  Ionization of He is ignored in this paper.
The gas density is written
\begin{equation}
\rho=(n_p+n_{\rm H})(1+4f_{\rm He}+\sum m_mf_m)m_p,
\end{equation}
where $m_m$ is the metal atomic mass in atomic units.

The pressure and the column density are integrated inward, with a starting value $P_{\rm top}=5\times 10^{-5}\mu$bar on the outside, where the atmosphere above becomes optically thin to \Lya. The starting value of each $N_i=0$. The solution is integrated inward from a starting radius $R_{\rm top}$ and pressure $P_{\rm top}$, until the base radius $R_b$ is reached. The boundary condition imposed there is that $P=P_b$. This boundary condition is satisfied in practice by varying the starting radius $R_{\rm top}$ until $P(R_b) = P_b$, the desired value, using Brent's method~\citep{Recipes}.

The hydrostatic model will be inaccurate near the outer boundary, as a number of physical effects have been neglected, such as: outflowing gas from the planet, interaction with the stellar wind, strong magnetic forces, radiation pressure, and stellar tidal forces. The region where these effects may be appreciable will be estimated in Section~\ref{sec:ad_cool}. However, in the region where this model shows the dominant absorption by H(2$\ell$), the density is so high that these effects are negligible. Hence the hydrostatic model is sufficient for the purposes of this study.

\subsection{Ionization State and Temperature}

At each level of the atmosphere, the pressure $P$ and columns $N_i$ are given by the boundary conditions or the integration of Equations \ref{eq:hb} and \ref{eq:Ni}. The temperature and particle densities must then be updated to continue the integration.
Since the gas is not in local thermodynamic equilibrium (LTE), these quantities must be determined by solving rate equilibrium equations for ionization/recombination, heating/cooling, a charge balance equation, and the equation of state. The equations used are as following. Terms related to metal species will be discussed in more detail in Section~\ref{metal}.

The charge balance equation is
\begin{equation}
n_p+\sum n_{\rm M_{\,II}} + 2\sum n_{\rm M_{\,III}}=n_e,
\label{eq:number}
\end{equation}
where $n_{\MII}$ and $n_{\rm M_{\,III}}$ are the number density of first ionized and second ionized metal species respectively.  Higher ionization states are ignored as their abundance would be negligible for the given ionizing flux and particle densities.

The hydrogen ionization and recombination balance equation is 
\begin{multline}
  (\alpha_Bn_e+k^{(\rm O)}_{\rm ion}n_{\rm O\,I})n_p=n_{1s}(\Gamma_{1s}+C_{1s \to \infty}^{(e)}n_e+k^{(\rm O)}_{\rm rec}n_{\rm O\,II})\\
  +C_{2 \to \infty}^{(e)}n_en_2+\Gamma_{2p}n_{2p}+\Gamma_{2s}n_{2s}+\sum_{m}\Gamma_{m}^{\rm 2nd}n_{m},
  \label{eq:Hp}
\end{multline}
where $\alpha_B$ is the case B recombination rate, which is a good approximation for region deeper than $3\times 10^{-3}~\rm \mu bar$ where the atmosphere is optically thick to Lyman continuum photons near the ionization threshold. $k^{(\rm O)}_{\rm ion}$ and $k^{(\rm O)}_{\rm rec}$ are  rates of whichoxygen ionizes and recombines through charge exchange with hydrogen respectively.  The n=2 state has separate contributions from H(2s) and H(2p) as
\begin{equation}
C_{2 \to k}^{(e)}n_en_2=C_{2s \to k}^{(e)}n_en_{2s}+C_{2p \to k}^{(e)}n_en_{2p}.
\end{equation}
The last term in Equation \ref{eq:Hp} represents ionization from the H(1s) state due to photoelectrons created by metal ionization. 
Hence high energy photoelectrons created through ionization of metal species can have the same secondary ionization effect as Equation~\ref{ke}.  The secondary ionization rate due to metal species is then
\begin{equation}
  \label{eq:2nd}
\Gamma_{m}^{\rm 2nd}=\int_{\nu_{\rm th}}^\infty \sigma_{m,\rm pi}(\nu)k_e \frac{F_\nu}{h\nu}e^{-\tau_{{\rm att},\nu}}\dd \nu,
\end{equation}
where $\sigma_{m,\rm pi}(\nu)$ is the metal photoionization cross section.
When evaluating the excited state H abundance in Equation \ref{eq:Hp}, the approximation in Equation~\ref{eq:6} is applied.

The heating and cooling balance equation is
\begin{align}
\label{eq:heat}
\Big[ \big(13.6\, &\text{eV}C_{1s \to \infty}^{(e)}+10.2\, \text{eV}C_{1s \to 2}^{(e)}\big)n_{1s}\nonumber\\
&+3.4\, \text{eV}C_{2 \to \infty}^{(e)}n_{2}+\sum_m \Lambda_mn_m+\Lambda_{ff}n_p\nonumber\\
&+\langle E_{rr}\rangle\alpha_Bn_p + k_{\rm B}T\sum_m\alpha_mn_m^{\rm ion}\Big]n_e\\
=n_{1s}&Q_{1s}(N_{1s})+Q_{2p}n_{2p}+Q_{2s}n_{2s}\nonumber\\
&+10.2\text{ eV}~ n_e n_2 C_{2 \to 1}^{(e)}+\sum_mQ_mn_m\nonumber,
\end{align}
where $\Lambda$ stands for cooling function.  \citet{AGN2} give the free free cooling rate
\begin{equation}
\Lambda_{ff}=1.85\times10^{-27} ~T^{1/2}(\ergcps),
\end{equation}
where $T$ is in units of Kelvin.
The mean kinetic energy of the recombining electrons is~\citep{Draine}
\begin{equation}
\langle E_{rr}\rangle=[0.684-0.0416\ln(T_4)]k_{\rm B}T,
\end{equation}
where $T_4=T/10^4\ \rm K$.
In Equation \ref{eq:heat}, the symbol $Q$ represents the photoelectric heating rate, per photoionization, corrected for the secondary ionization effect. \citet{Dalgarno} find that secondary electrons give rise to approximately the same number of ionizations as 1s$\rightarrow$2p excitations, so the heat deposited into the atmosphere by one photoelectron with energy $E$ is taken to be
\begin{equation}
\eta(E)=E-(13.6{\rm eV} + 10.2{\rm eV})k_e,
\end{equation}
where the second term represents the energy lost by the photoelectron to ionizations and \Lya\ excitations. 
Thus, the net photoelectric heating rate is
\begin{equation}
  Q =\int_{\nu_{\rm th}}^\infty \eta\sigma_{pi}(\nu)\frac{F_\nu}{h\nu}e^{-\tau_{{\rm att},\nu}}\dd \nu.
  \label{eq:7}
\end{equation}

Given $P$ and the columns $N_i$, Equations \ref{eq:P}, \ref{eq:number}, \ref{eq:Hp}, and \ref{eq:heat} give four algebraic equations to solve for $T$, $n_e$, $n_p$, and $n_{1s}$ at this level in the atmosphere. A globally convergent Newton's method \citep{Recipes} is applied to solve the set of equations.

\subsection{Radiative Cooling Due to Metal Species}\label{metal}

Although H and He are by far the most abundant elements, their electron-impact line cooling rates are heavily suppressed at temperatures $T \la 10^4\ \rm K$ due to the high excitation energies. Metal line cooling due to electron impact followed by radiative de-excitation is an important coolant, especially near the base of the atmosphere at $T \la 8000\ \rm K$. 
The ionization/recombination rate equilibrium equation is included to determine the relative abundance of each ionization state. Transitions yielding large cooling rates are chosen from abundant elements, and by striking a balance between low excitation energies, $\Delta E$,  and large radiative decay rate $A_{ul}$. Solar abundance is assumed \citep{Asplund}.  The elements considered are O, C, Mg, Si, S, Na, and K. Although Mg was not a priori expected to be abundant in the upper atmosphere due to condensation \citep{Visscher, KoskinenII}, Mg$_{\rm\,I}$ is in fact detected in HD 209458b \citep{Vidal-Madjar13} and marginally detected in \HD\ \citep{Cauley2016}.

The abundance of each ionization state is set by solving for rate equilibrium between ionization and recombination. Only neutral, first, and second-ionized atoms are included.

As a special case, rather than photoionization, collisional, and radiative recombination, the ionization state of oxygen is determined by charge exchange with hydrogen.  Considering that $n_{\rm H}\gtrsim 10^5~\rm cm^{-3}$ everywhere in the model, the charge exchange rates in the high-density limit in \citet{Draine} are applied.  The energy differences between three fine-structure levels of neutral oxygen are ignored because they are much smaller than $k_{\rm B}T$.

The rate equilibrium equations are
\begin{align}
  (\Gamma_{\MI}+C_{\MI,\infty}^{(e)}n_e+k^{(\rm O)}_{\rm ion}n_p)n_{\MI}&=(\alpha_{\MII}n_e+k^{(\rm O)}_{\rm rec}n_{\rm 1s})n_{\MII} \nonumber\\
  \Gamma_{\MII}n_{\MII}&=\alpha_{\rm M_{\,III}}n_en_{\rm M_{\,III}} \\
  n_{\MI}+n_{\MII}+n_{\rm M_{\,III}}&=(n_{\rm H}+n_p)f_m, \nonumber
\end{align}
where $C_{\MI,\infty}^{(e)}$ is the electron collisional ionization rate, which is only considered for Na and K atoms.
The collisional ionization of other metal species are ignored because of the much higher ionization potential.
The secondary ionization states of Na and K are ignored.  The photoionization rates of all species from \citet{Verner95, Verner96ion}. The Na and K collisional ionization rates are given by \citet{Lennon}. 

The rates of \citet{Pequignot1991} are used to describe the recombination of C and O, of \citet{Shull} for that of Mg, Si, and S, of \citet{Verner96recomb} and \citet{Landini} for that of Na and K respectively.

Because the ionization potentials of Mg\UI, Si\UI, Na\UI, and K\UI\ are smaller than the \Lya\ energy, both continuum and \Lya\ photons contribute to their photoionization and photoelectric heating rates. The metal species photoionization rate is 
\begin{multline}
\Gamma_{m,\rm pi}=\int_{\nu_{\rm th}}^\infty \sigma_{m,\rm pi}(\nu)\frac{F_\nu}{h\nu}e^{-\tau_{{\rm att},\nu}}\dd \nu \\
+ \frac{ 4\pi\sigma_{m,\rm pi}(\nu_{\rm Ly \alpha}) }{h\nu_{\rm Ly \alpha}}\int J_\nu\dd\nu,
\end{multline}
where the first integral excludes the stellar flux contribution near the \Lya\ line, and the metal species photoelectric heating rate
\begin{multline}
  Q_{m,\rm pi}=\int_{\nu_{\rm th}}^\infty \eta\sigma_{m,\rm pi}(\nu)\frac{F_\nu}{h\nu}e^{-\tau_{{\rm att},\nu}}\dd \nu \\
  +  4\pi \sigma_{m,\rm pi}(\nu_{\rm Ly \alpha})\frac{\nu_{\rm Ly \alpha}-\nu_{\rm th}} {\nu_{\rm Ly \alpha}}\int J_\nu\dd\nu,
  \label{eq:8}
\end{multline}
where $\sigma_{m,\rm pi}(\nu_{\rm Ly \alpha})$ is only nonzero for Mg\UI, Si\UI, Na\UI, and K\UI. The mean intensity $J_\nu$ in these formulas denotes the intensity in the vicinity of the \Lya\ line, found as a result of the resonant scattering calculation in Section \ref{MC}.

The metal line cooling rates require a model for the excited state densities.
For a two-level system, rate equilibrium between upward and downward rates gives
\begin{equation}
n_l(n_eC_{lu}^{(e)}+B_{lu}\bar{J}_{lu})=n_u(A_{ul}+n_eC_{ul}^{(e)}).
\end{equation}
Stimulated emission is ignored due to dilution of the stellar flux.
Collisional excitation is a sink of thermal translation energy while collisional de-excitation is a source.  Thus the cooling rate of this two levels system is
\begin{multline}
\Delta E\, n_e(C_{lu}^{(e)}n_l-C_{ul}^{(e)}n_u)\\
= n_e n_l \ \Delta E\ C_{lu}^{(e)}\left(\frac{A_{ul}-B_{lu}\bar{J}_{lu}(n_l/n_u)_{\rm eq}}{A_{ul}+n_eC_{ul}^{(e)}}\right)
\equiv \Lambda n_e n_l,
\label{eq:coolingfunc}
\end{multline}
where $(n_l/n_u)_{\rm eq}=g_l/g_u\exp(\Delta E/(k_{\rm B}T))$, and the last equality defines $\Lambda(T)$, the cooling function.  
Permitted transitions in the optical and near UV bands are always associated with strong absorption features in the stellar spectrum, which lead to a very small radiative excitation rate $B_{lu}\bar{J}_{lu}(n_l/n_u)_{\rm eq}$ for the transitions used.  In the case of forbidden transitions, to compensate for the small $A_{ul}$, only small $\Delta E$ transitions give rise to significant cooling.  At the long wavelength end, the dilution of radiation flux due to the solid angle of the star overcomes the effect of higher brightness temperature of the star.  As a result, radiative excitation is negligible in Equation \ref{eq:coolingfunc} for both permitted and forbidden transitions.   

For forbidden transitions, the electron number density $n_e$ is much larger than the critical density $n_{\rm crit}$ above which collisional de-excitation dominates radiative de-excitation. In this limit, the level population is given by the Boltzmann distribution, and the cooling rate becomes
\begin{equation}
\label{eq:11}
\Lambda n_e n_l=\Delta En_lA_{ul}\left(\frac{n_u}{n_l} \right)_{\rm eq},
\end{equation}
which is independent of $n_e$ and collisional rates.  

The emitted metal line photons are assumed to escape freely from the atmosphere.  In reality, the atmosphere may be optically thick to permitted emissions near the base of the atomic layer.

The major cooling processes are listed in Table~\ref{cooling}, while Table~\ref{tab:cooling_minor} in the appendix contains addition transitions from O, C,  S, and Si lines which are included in the model but only have a minor effect on the temperature. The lower state of some transitions may not be the ground state.   In this case, the lower state is assumed to reach collisional equilibrium with ground state in the cooling rate calculation.

Mg\UI\ 5184~\AA\ is another line that may contribute to cooling.  The lower state of this transition is not the ground state, and the upper state is associated with the ground state of the strong resonance line Mg\UI\ 2852~\AA.  
A Mg level population calculation and Mg\UI\ 2852~\AA\ radiation transfer study are required to accurately model the cooling effect, which is beyond the scope of this work.  Mg\UI\ 5184~\AA\ is not included in this work as a result.

\begin{table*}
\begin{center}
\caption{List of major metal cooling transitions}\label{cooling}
\begin{tabular}{cccccc}
\hline  \hline
Transition & $\Delta E$ & $A_{ul}$\footnote{\citet{NIST}} & $n_{\rm crit}$ \footnote{$n_{\rm crit}=A_{ul}/C_{ul}^{(e)}(4000\rm~K)$} & $\Lambda$  & Source of \\
 & (eV) & ($\rm s^{-1}$) & ($\rm cm^{-3}$) & ($\ergcps$) & collision rate \\
\hline
Mg\UI\ 2852 & 4.35 & $4.91\times 10^8$ & $6.7\times 10^{15}$ & $3.4\times 10^{-19}T^{0.18} e^{-5.04/T_4}\ $\footnote{$T_4=T/10^4{\rm~K}$} & Van Regemorter formula \footnote{\citet{vanRegemorter}}\\
Mg\UI\ 4571 & 2.712 & 254 & $8.5\times 10^9$ & $1.0\times 10^{-16} e^{-3.15/T_4}/(254+3.0\times 10^{-8}n_e)$ & \citet{Osorio}\\
Mg\UII\ 2803 & 4.422 & $2.57\times 10^8$ & $8.6\times 10^{14}$ & $7.1\times 10^{-12} C_{lu}^{(e)}$ & CHIANTI \footnote{~\citet{Chianti,Chianti8}}\\
Mg\UII\ 2796 & 4.434 & $2.60\times 10^8$ & $2.2\times 10^{14}$ & $7.1\times 10^{-12} C_{lu}^{(e)}$ & CHIANTI\\
Na\UI\ 5890+5896 & 2.104 & $6.16\times 10^7$ & $4.4\times 10^{14}$ & $3.4\times 10^{-12} C_{lu}^{(e)}$ & \citet{Igenbergs} \\
K\UI\ 7665+7699 & 1.615 & $3.78\times 10^7$ & $1.8\times 10^{14}$ &  $3.7\times 10^{-19}T^{0.18} e^{-1.87/T_4}$ & Van Regemorter formula \\
\hline
\end{tabular}
\end{center}
\end{table*}

\subsection{Molecular Hydrogen}\label{sec:H2}

Near the base of the atomic layer, where the temperature drops below $T \sim 2000-3000\ \rm K$, it is expected that the density of molecules will increase rapidly and come to dominate over the atomic species. This is the base of the atomic layer and the top of the molecular layer. The atomic-to-molecular transition is not self-consistently modeled in this work, as the rate equations to determine molecular densities (e.g. \citealt{Yelle,Garcia}), and the strong effect of molecular coolants from e.g. H$_2$O rotation-vibration bands, are not taken into account. 

Although the details of molecule formation are beyond the scope of this paper, a rough estimate of the H$_2$ number density is made to verify that the temperature does indeed become low enough to form molecules as the base is approached. \citet{Lenzuni} noted that for a wide range of radiation intensity, that dissociation of H$_2$ is due to collisional processes, rather than photo-dissociation. \citet{Yelle} found the same result for a model for the thermosphere of  HD 209458b. In this limit, an estimate of the H$_2$ density can be found using the Saha equation. Applying the H$_2$ partition function from \citet{Borysow}, the H$_2$ number density, $n_{\rm H_2}$ can be computed from the atomic hydrogen density, $n_{\rm H}$, and the temperature $T$.

The buildup of a significant column of H$_2$ will shield the lower atmosphere from stellar UV, and allow the formation of other molecules, e.g. CO and H$_2$O which may be important coolants. Another effect more relevant to this work is that H$_2$ may act as a strong absorber of \Lya\ photons, effectively setting a lower boundary to the region of the atmosphere that may have a large \Lya\ intensity. \citet{Black1987} discussed numerous ``accidental resonances" between \Lya\ and electronic transitions in H$_2$. The strongest of these can have oscillator strength $f \sim 10^{-2}$, implying that a column $N_{\rm H_2} \simeq 10^{14}\ \rm cm^{-2}$ is required to give unit optical depth for these transitions. For a scale height $H \simeq 10^8\ \rm cm$, this gives a critical number density $n_{\rm H_2} \simeq 10^6\ \rm cm^{-3}$ for true absorption optical depth unity over a scale height. In practice, \Lya\ photons near the optically-thick base of the atomic layer would require a large number of scatterings to escape the atmosphere, implying a total distance traversed even larger than a scale height. This will effectively set an absorbing lower boundary for the atomic layer in the radiation transfer described in Section~\ref{MC}.

\section{\Lya\ radiation transfer}\label{MC}

\Lya\ photons can excite hydrogen from the 1s to the 2p state, providing a population of absorbers that may be detected in Balmer line transmission spectra.  \Lya\ may also play a role in the heating/cooling and ionization/recombination balance, so a detailed \Lya\ radiation transfer calculation is crucial.

Two sources of \Lya\ are included in the model, the stellar \Lya\ incident through the top of the planet's atmosphere, and also \Lya\ generated within the planet's atmosphere by electron impact excitation or a recombination cascade. The results of particle densities and temperature versus radius from the hydrostatic model in Section \ref{hydro} are used to specify the \Lya\ source function, as well as the mean free paths to scattering and true absorption.
In the transfer calculation it is convenient to use \Lya\ line center optical depth, $\tau$, as the vertical coordinate.
A plot of pressure P and radius r versus $\tau$ will be shown in Figure~\ref{fig:T}.
A major simplifying assumption is to use plane-parallel geometry, so that mean intensity $J_\nu(r)$ is tabulated as a function of altitude.  The radius in the hydrostatic model varies by a factor of 2 from base to top, and by a factor of $\simeq20$\% near the base of the layer where the \Ha\ line is formed. The plane-parallel assumption simplifies the calculation, and is consistent with uniform irradiation assumed in the hydrostatic model.

At the outer boundary ($\tau=0$), the unpolarized stellar \Lya\ intensity enters the slab vertically. The line is parameterized by a double Gaussian line profile with width (in velocity units) $\sigma = 49\ {\rm km\ s^{-1}}$ and centers at $\mu = \pm 74\ {\rm km\ s^{-1}}$ \citep{Gladstone, Curdt, Tian}.  Instead of using the value of $\epsilon$ Eri from MUSCLES, the integrated line flux at the top of the \HD's atmosphere given by \citet{Linsky}, $F_0 = 2.0 \times 10^4\ergcs$, is applied.  Outgoing photons can escape from the top boundary freely.

An absorption bottom boundary is applied at the base of the slab, which represents the base of the atomic layer.  
Physically, this boundary condition is imposed to represent the short mean free path to true absorption of \Lya\ on H$_2$ \citep{Shull1978,Black1987} and H$_2$O \citep{Miguel}. Since $n_{\rm H_2}$ increases inward much faster than $n_{\rm H}$, the mean free path to resonant scattering will become longer than that to true absorption in the molecular layer, greatly decreasing the \Lya\ intensity compared to the atomic layer.  As will be shown, even in the atomic layer, the intensity falls rapidly toward the base. 

\Lya\ generated inside the atmosphere by collisional excitation or a recombination cascade is assumed to be unpolarized and have an initial frequency distribution given by a Gaussian distribution with the Doppler width set by the local temperature.  In practice, this initial distribution is much narrower than the resultant mean intensity, so it is the same effect as initializing photons at line center. 
The source function inside the atmosphere is
\begin{multline}
S_\nu=\frac{10.2{\rm~eV}}{4\pi}\Big(\alpha_Bn_pn_e + C_{1s\to 2}^{(e)}n_{1s}n_e\\
+\Gamma_{1s}^{\rm 2nd}n_{1s}+\sum_m \Gamma_{m}^{\rm 2nd}n_m\Big)\phi_\nu,
\label{eq:sourcefunc}
\end{multline}
where $\phi_\nu$ is the Doppler profile evaluated at the local temperature. The first term in Equation \ref{eq:sourcefunc} represents recombinations, each of which is assumed to produce one \Lya\ photon. The following three terms represent collisional excitation by thermal electrons, photoelectrons from ionizing H(1s), and photoelectrons generated from ionizing metals.  Recombination and collisional excitation to H(2s) are also included because H(2s) and H(2p) are well coupled by the $\ell$-mixing.

The line center optical depth to scattering reaches values as large as $\tau \simeq 10^8$ near the base of the model, and photons will scatter $\sim \tau$ times before exiting the atmosphere \citep{Harrington} if no other process intervenes. \Lya\ photons can leave this scattering cycle by two categories of processes. First, the radiative excitation may not be followed by radiative de-excitation of a \Lya\ photon some fraction $\epsilon$ of the time. 
The dominant processes are: photoionization from the n=2 state by stellar Balmer continuum emission; collisional de-excitation; and two photon decay from 2s$\rightarrow$1s.  The former two processes also contribute to ionizing and heating rates.  The second kind of process is ``true absorption", in which a \Lya\ photon is absorbed by some other species besides H(1s), for example by photoionizing an atom with low ionization potential less than 10.2eV.

\Lya\ photons can also leave this scattering cycle by the $\ell$-mixing or radiative excitation processes from the 2p state.  Because it will be followed by the reverse process immediately, these processes are equivalent to \Lya\ photon redistribution that put the line wing photons frequently back to the line center and stop the photons from escaping.
These processes are not included in this model because a careful consideration required the \Ha\ mean intensity in the atmosphere, which is not available right now.

The plane-parallel transfer equation including resonant scattering, true absorption, the source function for photon creation, and excitation followed by de-excitation a fraction $1-\epsilon$ of the time is
\begin{multline}
\mu\frac{\dd I_\nu(z,{\bm n})}{\dd z}=-(\alpha_{\rm sc}+\alpha_{\rm abs})I_\nu(z,{\bm n})+ S_\nu(z) \\
+(1-\epsilon)4\pi \frac{\alpha_{\rm sc}}{\phi_\nu}\int R(\nu,{\bm n};\nu',{\bm n'}) I_{\nu'}(z,{\bm n'}) \dd \Omega' \dd \nu',
\end{multline}
where $\mu=\cos\theta$, $\phi_\nu$ is the Voigt line profile, $\alpha_{\rm sc}$ is resonant scattering coefficient, $\alpha_{\rm abs}$ is true absorption coefficient, and $R$ is the \citet{Hummer} case II-B redistribution function, with dipole angular dependence. 
The redistribution function $R$ gives the probability that the photon is scattered from incident frequency $\nu'$ and direction ${\bm n'}$ to frequency $\nu$ and direction ${\bm n}$. Case II-B redistribution assumes that, in the rest frame of the atom, the line profile for absorption of the photon (excitation) is a Lorentzian profile, and that in the rest frame the outgoing photon has the same energy as the ingoing photon. \citet{Hummer} presents formulae for the resulting redistribution function thermally averaged over a Boltzmann distribution of atom velocities. Case II-B with dipole angular dependence results for resonant scattering when the initial and final states are the H(1s) state, and the finite lifetime of the intermediate H(2p) state is included. Fine structure splitting of the excited state is ignored. This is a good approximation in the present application where the mean intensity is much broader than the fine structure separation of the $J=1/2$ and $3/2$ states.

The transfer equation is solved numerically with the Monte Carlo method (e.g. \citealt{2011BASI...39..101W}).  
First, unpolarized photon packets are initialized at a point randomly generated from the source function (see Equation~\ref{eq:sourcefunc}) and with a randomly chosen direction.
Second, the optical depth $\tau$ that the photon will travel through before it is scattered or absorbed is randomly generated with a probability density $e^{-\tau}$.  The spatial location of the scattering or ``true absorption'' at optical depth $\tau$ from the point of emission is then determined with the knowledge of the densities and temperature, and cross sections of $n_{\rm H}$, $n_{\rm Mg I}$, $n_{\rm Si I}$, $n_{\rm Na I}$, $n_{\rm KI}$.  
The Lucy method~\citep{Lucy} is used to tabulate intensity and flux from the motion of the photon packets.  This was crucial in optically thick regions, and worked much better than accumulating photon statistics only when they pass through the face of a cell.  
Based on the radiative excitation and absorption optical depth, the rejection method is used to determine whether the photon is absorbed.  Whether the H(2p) emits another \Lya\ photon is determined by comparing a random number with $1-\epsilon$.  
Third, at each scattering, the outgoing photon direction is chosen including polarization accumulated during prior scatterings.  The Stokes matrix for Rayleigh scattering is used.  However, rather than using the Stokes matrix in scattering-plane coordinates, as discussed in ~\citet{Chandra}, a scattering matrix was derived in terms of the incoming and outgoing photon direction without reference to the scattering plane, which was found to be more convenient for numerical calculations.  
Then, the velocity of scatterers along the direction of the incident photon is randomly generated by the method described in \citet{Zheng}, with small modifications to improve the efficiency.  Recoil is included in computing the new frequency of the photon after scattering.  The process of propagation and scattering is repeated until the photon escapes the modeled system or leaves the scattering cycle.

By knowing the $n_{1s}$ from the hydrostatic atmosphere model and $\bar{J}_{Ly\alpha}$ from the \Lya\ radiation transfer calculation, the $n_{2\ell}$ population is given by Equation~\ref{eq:6}.
Note that the ionization, heating, and cooling rates in the atmosphere depend on the $n_{2\ell}$ and \Lya\ mean intensity.
Therefore, the hydrostatic model and radiation transfer calculation are performed iteratively, until the $n_{2\ell}$ converged, which takes typically $\sim 8$ iterations.

\section{the fiducial atmosphere model}
\label{sec:fiducial}

\begin{figure}
  \centering
  \includegraphics[width=0.5\textwidth]{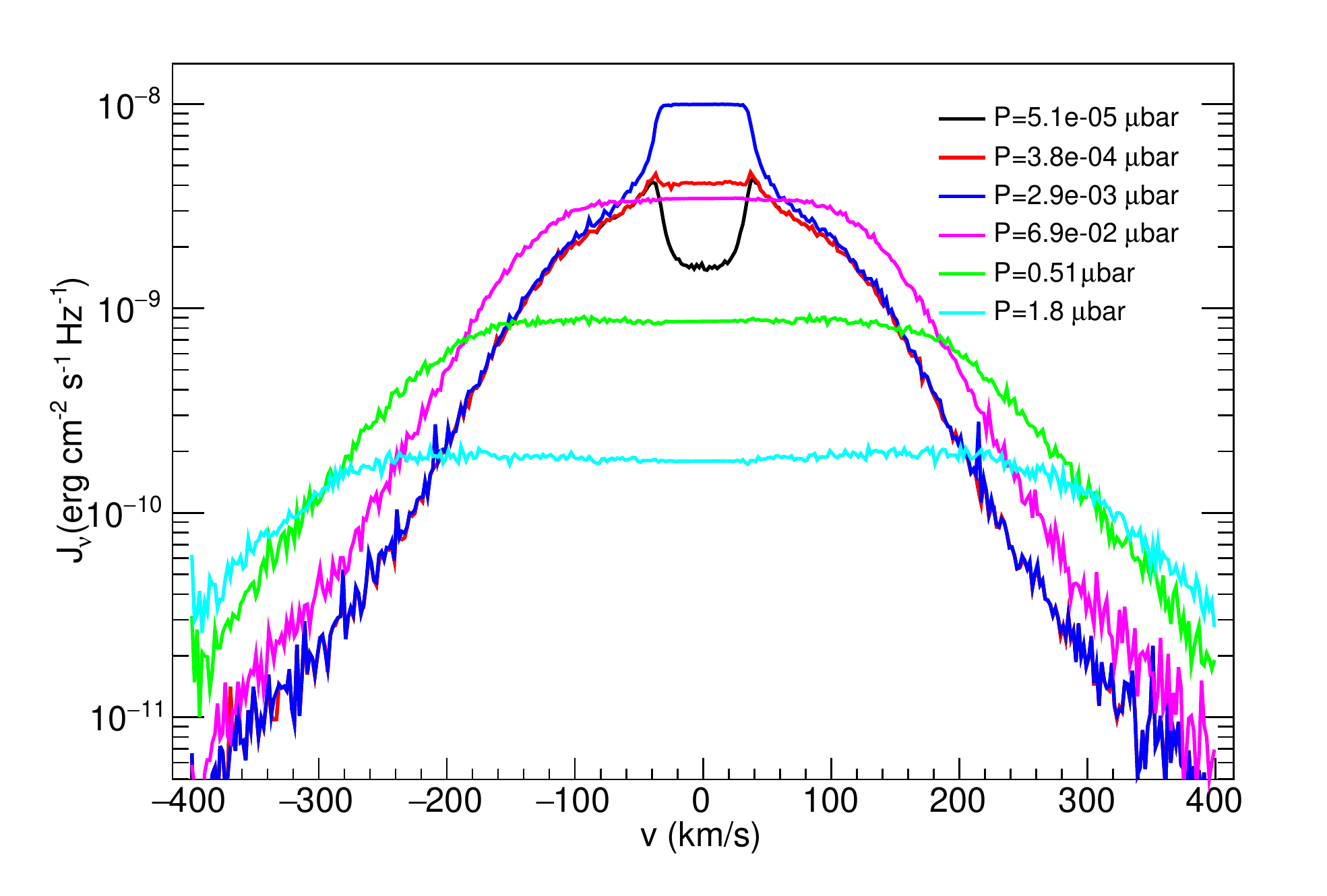}
  \caption[mean intensity]{\Lya\ mean intensity spectrum $(J_{\nu})$ at six different depths in the atmosphere.  At line center the spectrum is flat, much wider than the thermal line width, and becomes broader with depth.  The intensity decreases rapidly on the line wing.  The fluctuations in the spectra are due to the statistical noise in the Monte Carlo simulation.}
  \label{fig:Jnu}
\end{figure}

\begin{figure}
  \centering
  \includegraphics[width=0.5\textwidth]{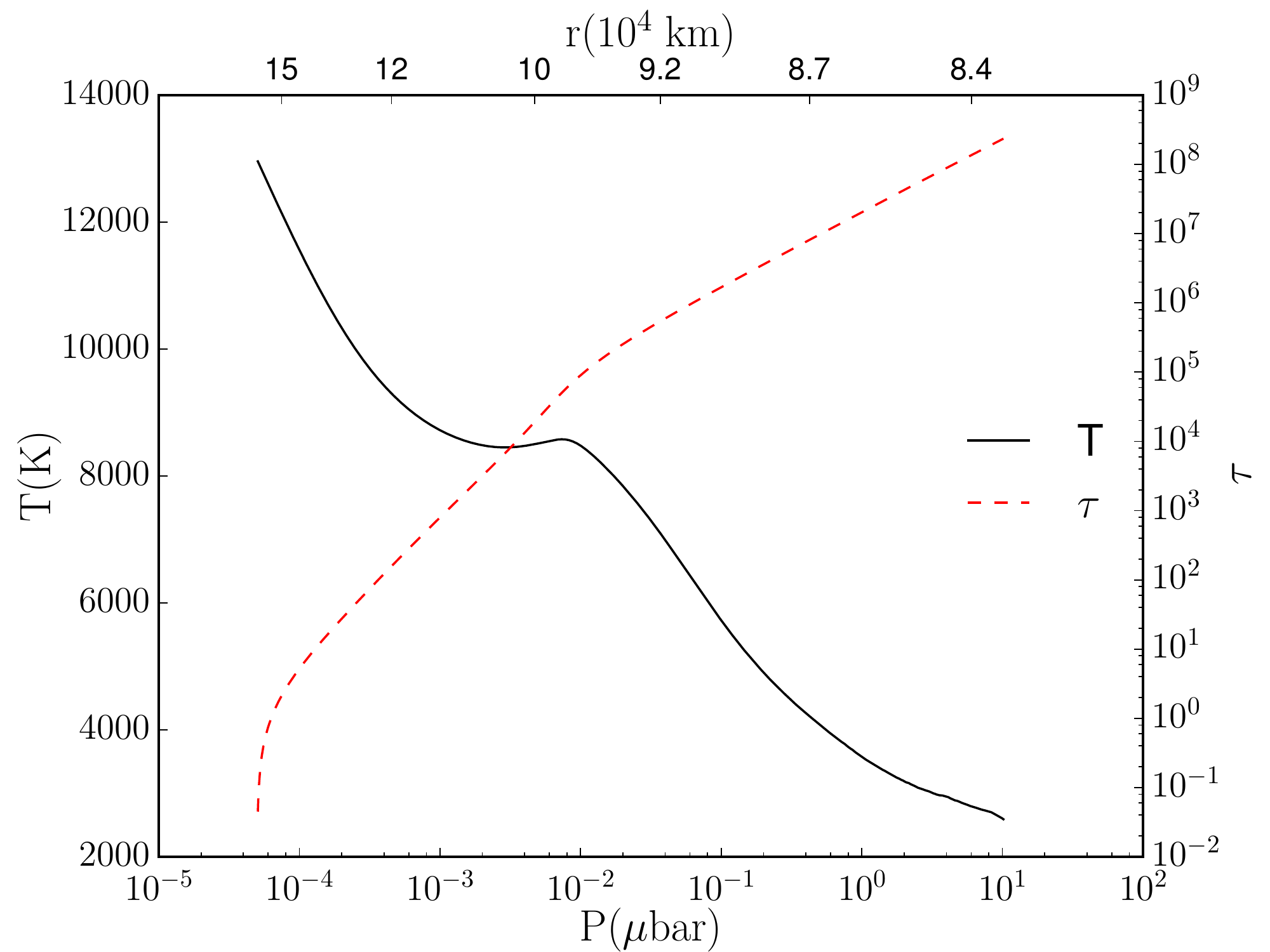}
  \caption[Temperature]{Temperature ($T$) and \Lya\ line center optical depth ($\tau$) versus pressure ($P$, bottom axis) and radius ($r$, top axis).} 
  \label{fig:T}
\end{figure}

\begin{figure*}
  \centering
  \includegraphics[width=\textwidth]{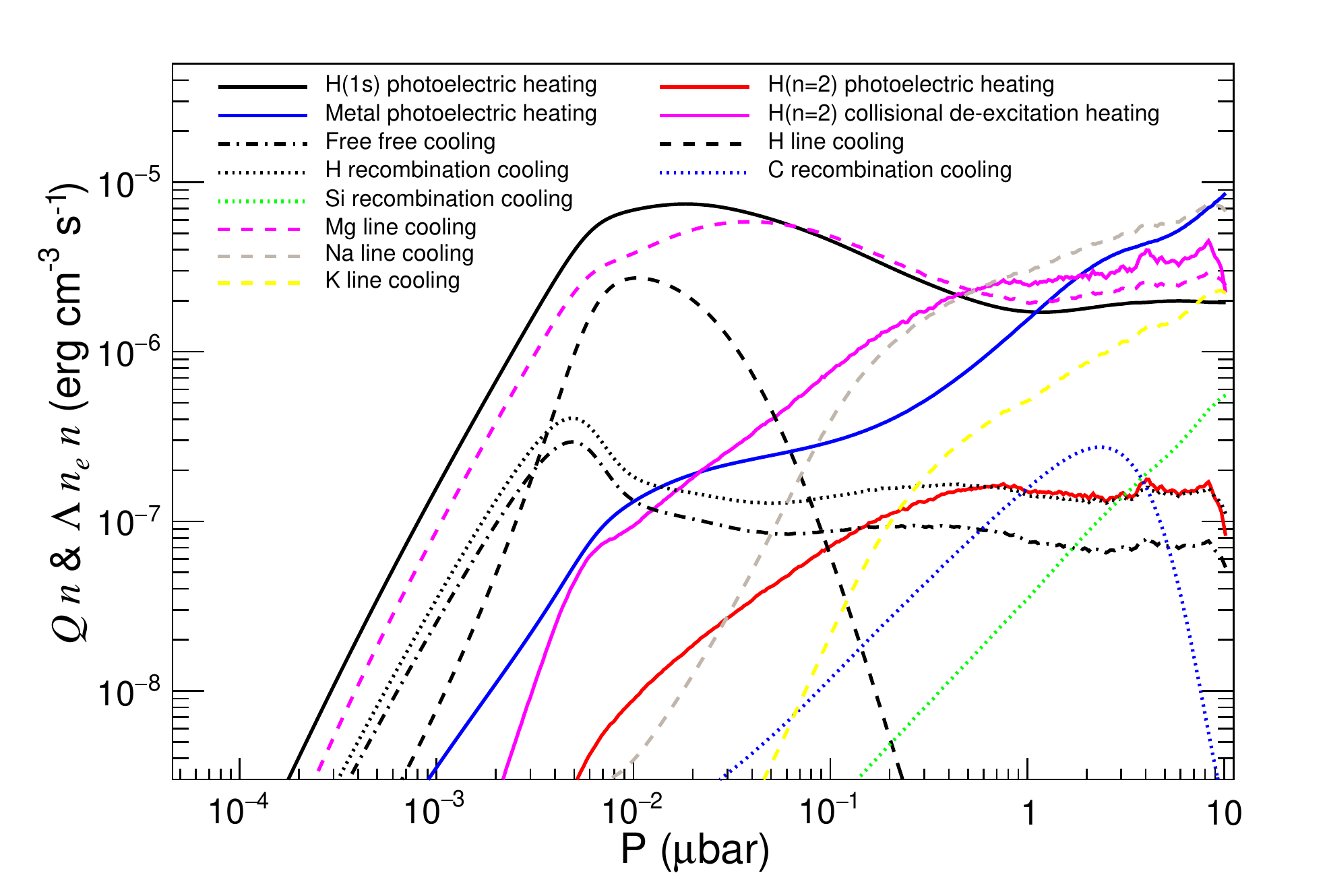}
  \caption[Cooling rate]{Major heating ($Qn$, solid lines) and cooling ($\Lambda n_e n$) rates versus pressure ($P$). The line cooling profiles (dashed lines) present the radiative cooling of \Lya\ and the metal lines listed in Table~\ref{cooling}.}
  \label{fig:cooling}
\end{figure*}

\begin{figure*}
  \centering
  \includegraphics[width=\textwidth]{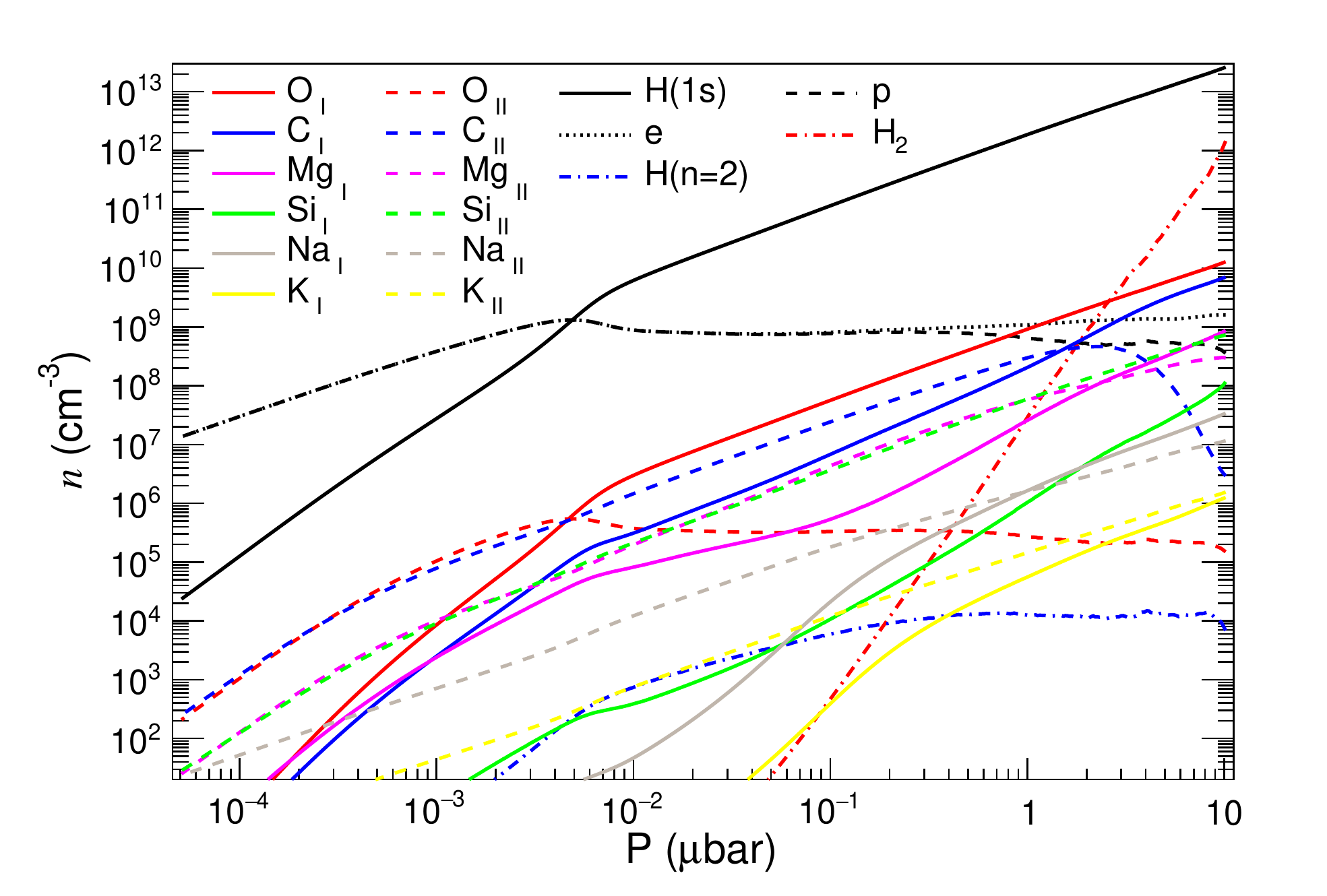}
  \caption[number density]{Number density ($n$) of main species against the pressure ($P$).  The solid lines present the profiles for neutral atoms and the dashed lines for first ionized ions.  The combination of large \Lya\ intensity $\bar{J}_{Ly\alpha} \propto P^{-1}$ and increasing H(1s) density with depth give rise to an approximately flat H(2$\ell$) density around $10^4~\rm cm^{-3}$ over two decades in pressure near the base of thermosphere.  The number densities of Na\UI\ and H(2$\ell$) are comparable in the pressure region between $10^{-2} \mu$bar and $10^{-1} \mu$bar, where the line center optical depths of \Ha\ and Na D are near unity (see Figure~\ref{tau_b}). }
  \label{fig:n_P}
\end{figure*}

\begin{figure*}
  \centering
  \includegraphics[width=\textwidth]{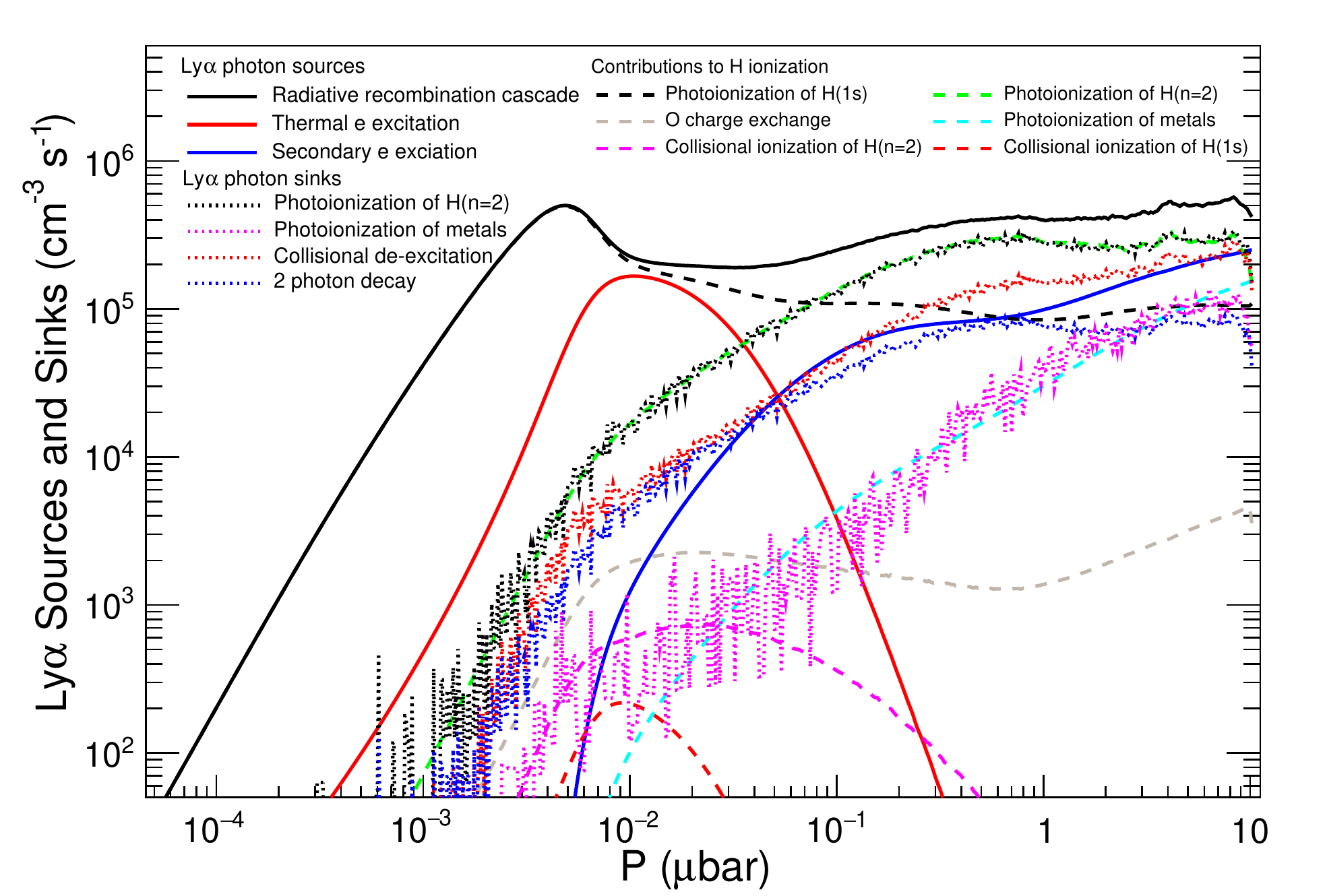}
  \caption[Source&Sink]{\Lya\ photon sources, sinks, and H ionization rates per unit volume as a function of pressure ($P$).  Each \Lya\ photon source in Equation~\ref{eq:sourcefunc} is plotted with a solid line.  The secondary $e$ excitation stands for the collisional excitation by $e$ generated by the photoionization of H($1s$) ($\Gamma_{1s}^{\rm 2nd}n_{1s}$) and metals ($\sum_m \Gamma_{m}^{\rm 2nd}n_m$).  The sink rates are the output from the \Lya\ Monte Carlo simulation.  The process that a \Lya\ photon photoionizes a low ionization potential metal atom is referred as the photoionization of metals under the \Lya\ photon sinks section.  A breakdown of the H radiative recombination cascade rate into individual H ionization processes based on Equation~\ref{eq:Hp} is also plotted.  The output rate of photoionization of H(2$\ell$) in the \Lya\ Monte Carlo simulation as a photon sink recovers the rate of the same process as a H ionization given by the ionization equation.  O charge exchange stands for the difference between recombination of an O\UII\ and ionization an O\UI\ by charge exchange with H ($k^{(\rm O)}_{\rm rec}n_{\rm O\,II}n_{1s}-k^{(\rm O)}_{\rm ion}n_{\rm O\,I}n_p$).  The photoionization of metals listed here stands for the H collisional ionization by $e$ generated by the photoionization of metals ($\sum_m \Gamma_{m}^{\rm 2nd}n_m$). }
  \label{fig:source}
\end{figure*}

\begin{figure}
  \centering
  \includegraphics[width=0.5\textwidth]{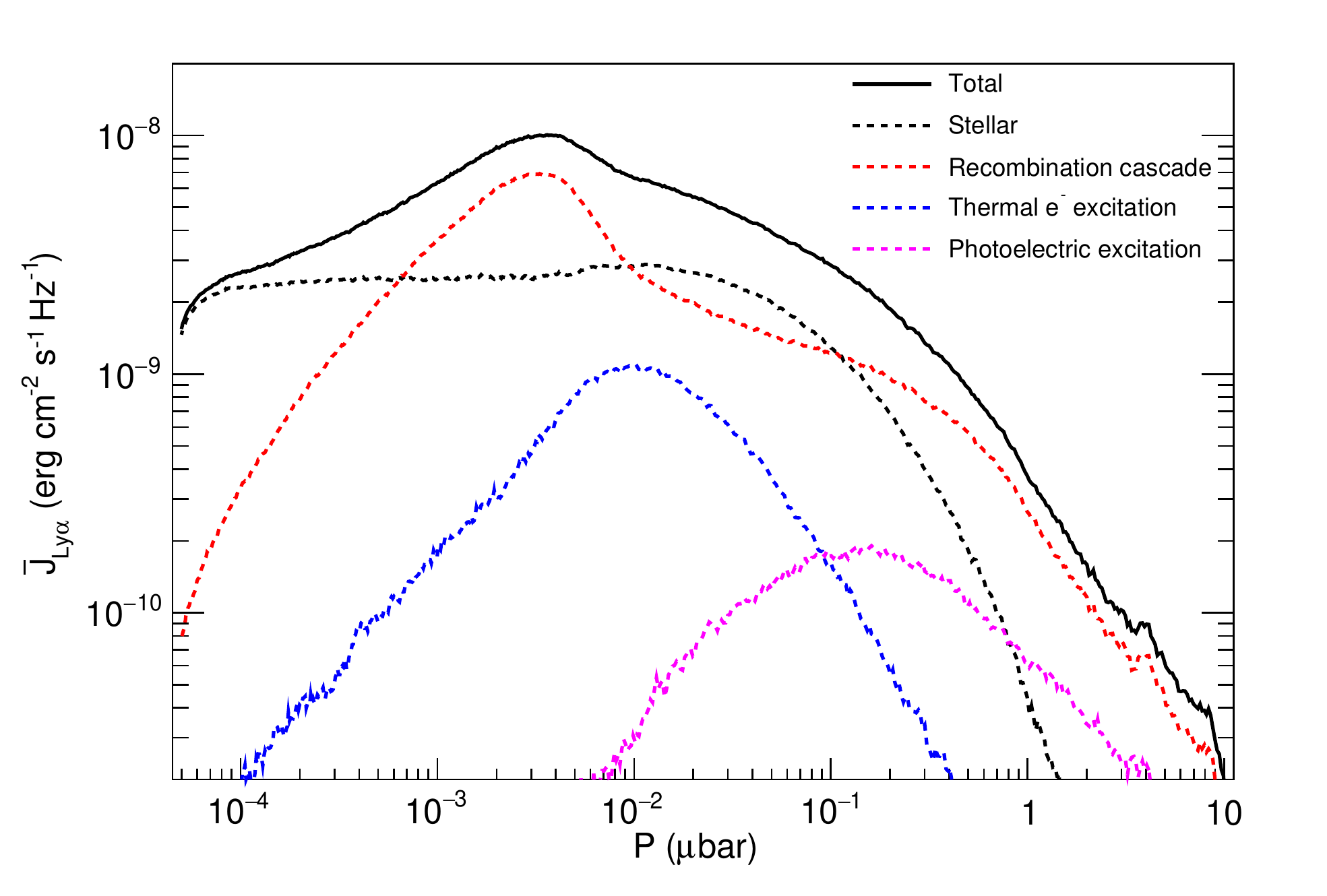}
  \caption[weighted mean intensity]{Line profile weighted mean intensity $\bar{J}_{Ly\alpha}$ from different initial \Lya\ photon generation mechanisms against the pressure ($P$).  The large $\bar{J}_{Ly\alpha}$ stays nearly constant deep into the atmosphere and $\bar{J}_{Ly\alpha} \propto P^{-1}$ near the base of the atomic layer.  Both \Lya\ photons created inside the atmosphere and incident from the star are important in the \Ha\ line formation region. }
  \label{fig:Jbar}
\end{figure}

In this section, results are presented for the fiducial model with solar abundance for all species and extreme ultraviolet (EUV) and X-ray flux set by the day-side value.  The \Lya\ mean intensity spectrum in the atmosphere is shown in Figure~\ref{fig:Jnu}.  Temperature versus height for the fiducial model is shown in Figure~\ref{fig:T}. The heating and cooling rates per unit volume are given in Figure~\ref{fig:cooling}, and the number densities of each species are given in Figure~\ref{fig:n_P}.  Figure~\ref{fig:source} shows the \Lya\ photon sources, sinks, and H ionization rates per unit volume.  Line profile weighted \Lya\ mean intensity $\bar{J}_{Ly\alpha}$ of the fiducial model are given in Figure~\ref{fig:Jbar}.

Figure~\ref{fig:Jnu} shows the \Lya\ mean intensity at six different depths in the atmosphere.  The line center of the resonant scattering spectrum has a flat part with width $\sim (a\tau)^{1/3}\nu_D$~\citep{Harrington}.  The dip at line center near the surface arises because the photons have to diffuse away from line center in order to escape from the slab, due to the extremely short mean free path at line center.  The mean intensity near line center has a peak around $\tau\sim 10^4$, where the H($1s$) photoionization, as well as the subsequent recombination, which is the major \Lya\ photon source, is the strongest.  The spectrum decreases rapidly on the line wing.  The fluctuations in the spectra are due to the statistical noise in the Monte Carlo simulation.  Since the number densities of H(2$\ell$) are proportional to the $\bar{J}_{Ly\alpha}$, this fluctuation also appears in the H(2$\ell$) number densities curve, and other curves such as heating rates. 

At the top of the model, for pressures $P<3\times 10^{-3}\, \mu$bar, the gas is fully ionized and the contribution to the number density of electrons by metal species is negligible.  The gas is optically thin to LyC photons, thus the ionization rate and heating rate per particle are nearly constant.  In this region $n_{1s}\approx n_e^2\alpha_B/\Gamma_{1s}\propto P^2$.  The gas temperature is set by the balance of H($1s$) photoelectric heating and line cooling by \ion{Mg}{2} for solar Mg abundance or \Lya\ for low Mg abundance.  The cooling rate $n_{\rm Mg II}n_e\Lambda_{\rm Mg}\approx n_{1s}Q_{1s}\propto P^2$.  Because the $n_{\rm Mg II}$ increases faster than $P$ (see Figure~\ref{fig:n_P}), the gas temperature is regulated to $T\propto 1/\log P$ near temperatures $T=9000 \-- 13,000$ K, shown in Figure~\ref{fig:T}.

Note that the ionized region of the hydrostatic model at $P\lesssim 3\times 10^{-4}\, \mu$bar may have an unphysically high temperature, as the inclusion of a hydrodynamic outflow and adiabatic cooling may be important in this region, as in the well-studied case of HD 209458b (e.g. \citealt{Yelle}). However, inspection of Figure \ref{fig:n_P} shows that the H(2$\ell$) and Na\UI\ densities are negligible in this region, and hence errors in the temperature profile there will not affect the \Ha\ and Na transit depth. 

H(1s) is the main absorber of the stellar LyC flux over the majority of the energy range.  Besides H(1s), the 2s-shell of O\UI\ absorbs most photons with energy above 538 eV, and C and Si are the main absorbers of the photons below 13.6 eV in the atomic layer.
The atmosphere becomes optically thick to 400\AArm\ photons at the pressure of $\sim 5 \times 10^{-3}~\rm \mu bar$.
The strong stellar flux between 100 and 400\AArm\ causes the local maximum in the $T-P$ profile.
Photoelectric heating from ionization of H(1s) contributes the large peak in Figure \ref{fig:cooling} over the pressure range $P=10^{-3}-10^{-1}\, \mu$bar. Below that, it continues to be an important source of heating, with absorption of successively higher energy photons with depth, and ionization by secondary electrons becoming important. There is a narrow region near $P\sim 1\, \mu$bar where heating due to electron impact de-excitation of H(2$\ell$) dominates. Ultimately the heating in this region is due to the \Lya\ radiation, which excites the atoms to the n=2 state. Near the base, at $P=1-10\, \mu$bar, metal photoelectric heating from ionization of Si, O, and C dominates the heating.
Na line cooling is the dominant coolant below $P=0.5\, \mu$bar.  Above that, assuming solar Mg abundance, Mg line cooling dominates, among which \ion{Mg}{1} 4571 contributes most at $P>0.4~\rm \mu bar $ while \ion{Mg}{2} contributes most above.  \Lya\ line would be the major coolant instead if the Mg abundance is low.

Hence near the base of the model, both heating and cooling are controlled by metal species, and the temperature is not sensitive to an overall shift in metallicity. Above $P=1\, \mu$bar, the temperature is sensitive to metallicity, and eliminating a single important coolant could make a difference, with Mg and Na-poor atmospheres expected to be hotter and more extended. Since this region is important for the \Ha\, and Na transmission spectrum, the temperature, with it's dependence on the metallicity, may be an important parameter in understanding the transmission spectrum.

\Lya\ line center optical depth versus height for the fiducial model is also shown in Figure~\ref{fig:T}. Near the base of the atomic layer, $\tau \propto P \propto n(H(1s))$, since ground state hydrogen is the dominant species. However, above $P=2\times 10^{-3}\, \mu$bar, hydrogen is predominantly ionized, and the optical depth decreases outward approximately as $\tau \propto n(H(1s)) \propto n_p^2 \propto P^2$ in ionization equilibrium. At the outer boundary of the model, $\tau$ drops abruptly to zero due to the $\tau=0$ boundary condition there.

The most abundant species in Figure \ref{fig:n_P} are electrons and protons above $P=5\times 10^{-3}\, \mu$bar, and H(1s) below.  Recall that ionization of He is ignored in this paper, and that He is assumed to be neutral and have solar abundance.
The electron number density stays nearly constant in the deeper part of the atomic layer because both the ionization and the recombination rates are insensitive to altitude and temperature.  The ionization is dominated by the photoionization from H(2$\ell$) in this region and a flat H(2$\ell$) number densities lead to a flat ionization rate (see Figure~\ref{fig:source}).  
From the near equality $n_e \simeq n_p$, ionization of hydrogen supplies most of the electrons down to $1\, \mu$bar, and first ionized Mg and Si are important below. The ionization state of O closely follows that of H because of the very large O and H charge exchange rate.  
The atmosphere becomes opaque to the stellar flux above 10.4 eV due to the absorption by S\UI\ and C\UI.  The atmosphere is transparent to stellar flux below the S\UI\ ionization threshold throughout the model.  
The ionization of Na\UI\ and K\UI\ is dominated by collisional ionization at the level above $0.1 \rm~\mu bar$.  

Near the base of the model, the density of H$_2$ rises rapidly, and is only slightly less abundant than H(1s).  In a more complete model it is expected that the inclusion of strong molecular cooling due to e.g. H$_3^+$ and H$_2$O would cause the temperature near the base to be even lower and $n_{\rm H_2}$ to be even larger.

The combination of large \Lya\ intensity $\bar{J}_{Ly\alpha} \propto P^{-1}$ (see Figure~\ref{fig:Jbar}) and increasing H(1s) density with depth gives rise to an approximately flat H(2$\ell$) densities around $10^4~\rm cm^{-3}$ over two decades in pressure near the base of atomic layer.  

The number densities of Na\UI\ and H(2$\ell$) are similar in the pressure region between $10^{-2} \mu$bar and $10^{-1} \mu$bar.  In Section \ref{sec:transit}, it will be shown that $\tau \sim 1$ for Na D and \Ha\ in this pressure region.  This leads to similar transit depths for the \Ha\ and Na D transmission lines, in agreement with the observations of \citet{Cauley2016} and \citet{Wyttenbach}.

Shown in Figure~\ref{fig:source}, the photoionization rate from H($1s$) dominates the H ionization rate in the top layer of the atmosphere, and becomes constant after the atmosphere becomes optically thick to the LyC photons.  The photoionization from H(2$\ell$) takes over for the region $P\gtrsim10^{-1}~\rm \mu bar$.  The rate of charge exchange between O and H can be very high, but they almost cancel each other and leave a small net effect.  The collisional ionization by secondary $e$ generated by the photoionization of metals ($\sum_m \Gamma_{m}^{\rm 2nd}n_m$) becomes large at $P>1~\rm \mu bar$.  

The H radiative recombination cascade process dominates the \Lya\ photon production throughout the simulation domain.  The radiative decay after a thermal $e$ collisional excitation from $1s$ state has a narrow peak in creating \Lya\ photon near $10^{-2}~\rm \mu bar$.  The collisional excitation by $e$ generated by the photoionization of H($1s$) ($\Gamma_{1s}^{\rm 2nd}n_{1s}$) and metals ($\sum_m \Gamma_{m}^{\rm 2nd}n_m$) becomes important in creating \Lya\ photons near the top and base of the atomic layer respectively.

The stellar \Lya\ photons incident through the surface and the photons generated above $10^{-2}~\rm \mu bar$ can mostly escape through the top boundary.  In contrast, the photons emitted below $0.1~\rm \mu bar$ are mostly absorbed during the resonant scattering processes due to the high optical depth.  The \Lya\ flux at the bottom boundary is about $40\ergcs$. 

The total LyC flux absorbed inside the simulation domain, is $F_{\rm LyC}=2.6 \times 10^4 \ergcs$.  Besides the $2.0 \times 10^4 \ergcs$ stellar \Lya\ flux that mostly reflects back out, a net $9.6\times 10^3 \ergcs$ flux leaving the atmosphere originated from the \Lya\ emission inside the atmosphere.  In comparison, integrating the source function in Equation~\ref{eq:sourcefunc} over height, solid angle and frequency, the total column \Lya\ internal emission is $1.6 \times 10^4 \ergcs$.  And the flux of \Lya\ that hits the bottom boundary is $44 \ergcs$.

Because the photoionization of an H(2$\ell$) is followed by a radiative recombination cascade,  emitting another \Lya\ photon, these two processes taken together can be thought of not as sources or sinks, but as a redistribution in photon energy. The absorption of a photon on the wing, and its subsequent re-emission at line center make it harder for the photon to escape the atmosphere. The most important remaining ``real'' photon sources are excitation by secondary $e$, and radiative recombination cascade of a $p$ which was collisionally ionized by high energy photoelectron.  The ``real'' photon sinks are collisional de-excitation, photoionization of metals by \Lya, and 2 photon decay.

A breakdown of line profile weighted mean intensity, $\bar{J}_{Ly\alpha}$, in terms of the different sources is given in Figure~\ref{fig:Jbar}.  The external \Lya\ directly from star stays nearly constant above $0.05~\rm \mu bar$.  Because the incident stellar \Lya\ is peaked at the frequency that is more than 5 Doppler widths away from the line center, the cross section to these photons in the Lorentzian wing is about $10^5$ smaller than the cross section in the line center.  Most stellar \Lya\ photons can directly penetrate to the $P\sim 10^{-2}\mu$bar level, which leads to a nearly constant intensity above this layer.
Radiative recombination cascades are important at $10^{-3} \-- 10^{-2}~\rm \mu bar$ and below 0.1 $\mu$bar, with secondary $e$ excitation becoming the second largest source deeper than $1~\rm \mu bar$.

The \Lya\ mean intensity near the base can be estimated by assuming a local balance of frequency-integrated sources and sinks, giving $J \simeq S L$. Here $J$ is the frequency integrated mean intensity, $S$ is the frequency integrated source function, and $L$ is the total path length traversed by the photon before it is destroyed. The photon source is insensitive to altitude as shown in Figure~\ref{fig:source}. 
For a sink given by true absorption due to photoionization of metals, the mean intensity is 
\begin{equation}
  \label{eq:2}
  J \simeq S L = \frac{S} {n_m \sigma_{m,\rm pi}(\nu_{Ly\alpha})}.
\end{equation}
The photoionization cross section of metals by \Lya\ photons, $\sigma_{m,\rm pi}(\nu_{Ly\alpha})$, is independent of altitude and frequency.  The number densities of Mg\UI, Si\UI, Na\UI, and K\UI\ scale proportional or slightly steeper than $P$.  As a result, the $J \propto P^{-1}$.
Next consider collisional de-excitation and 2 photon decay, for which
\begin{align}
  \label{eq:3}
  J &\simeq  S \epsilon^{-1} l_{\rm mfp} \nonumber \\
&= S \left(\frac{3 A_{2p\to 1s}}{4 n_eC_{2\to 1s}(T)+A_{2s\to 1s}}\right)\left(\frac{1}{n_{1s}\sigma_{1s}(\nu_0)}\right),
\end{align}
where $\epsilon$ is the probability of excitation not followed by de-excitation,  and $l_{\rm mfp}$ is the line center mean free path.  Because $n_e$, $C_{2\to 1s}(T)$ are insensitive to altitude,  $J \propto P^{-1}$ in this case also.  

\begin{figure}
  \centering
  \includegraphics[width=0.5\textwidth]{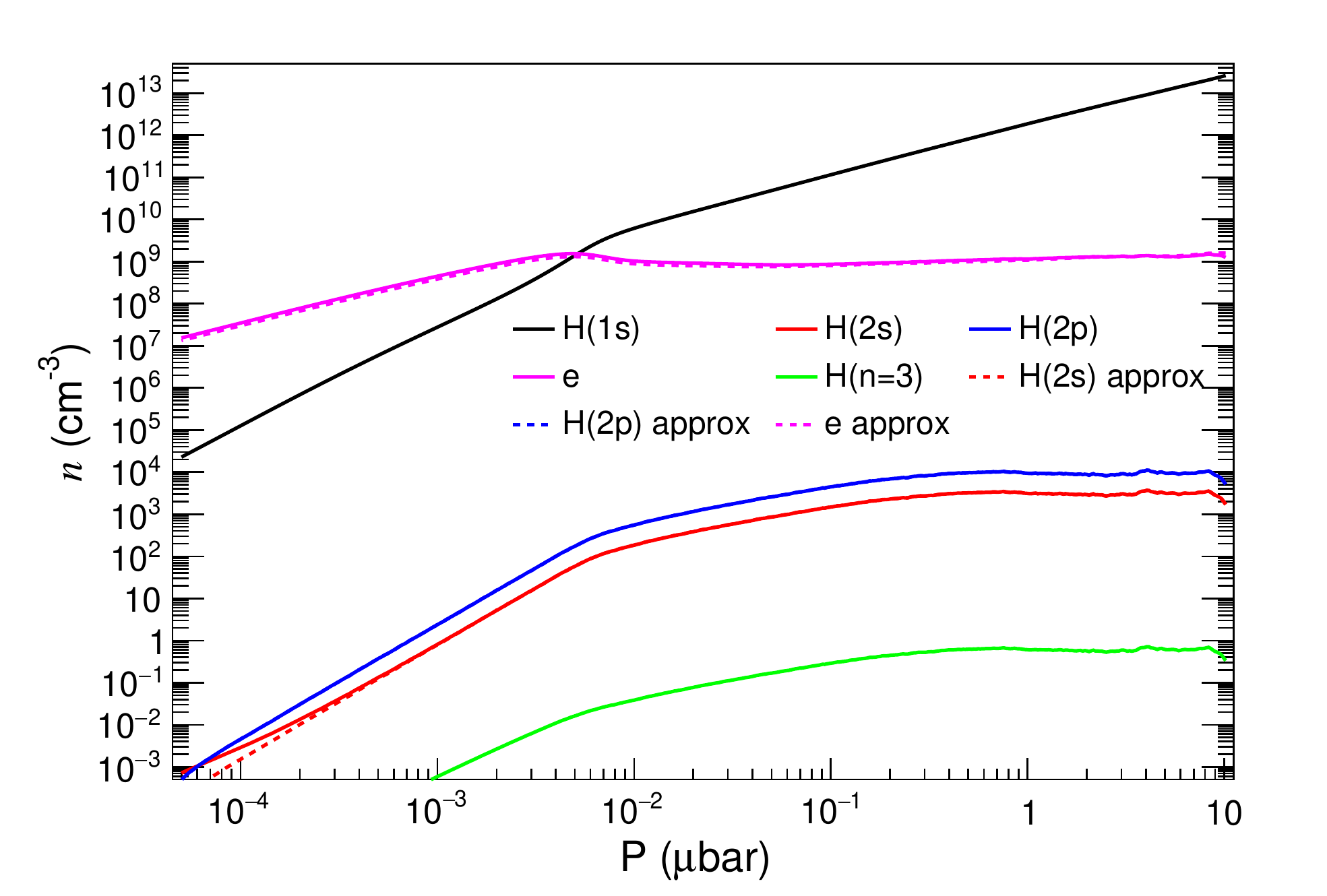}
  \caption{Comparison of the hydrogen sub-states and electron number densities ($n$) obtained from the full hydrogen level population code (solid line) and the approximation in Equation~\ref{eq:6} (dashed line), which mostly overlaps with the solid lines.  A rough estimate of the H($n=3$) number density is also shown in green.}
  \label{fig:pop}
\end{figure}

To show the reliability of the approximations given in Equation~\ref{eq:6}, which are used in the hydrostatic model, the results for the H($1s$) number density and temperature were plugged back to the full hydrogen level population code described in Section~\ref{sec:level}. Figure~\ref{fig:pop} compares the approximate result and the full calculation for the $n_e$, $n_{2s}$, and $n_{2p}$. The approximation holds for the whole simulation region.  The $n_{2s}$ and $n_{2p}$ obtained from both methods are almost identical except the approximate method slightly underestimates the $n_{2s}$ at the very top of the atmosphere, due to the fact that the contribution from recombination and collisional excitation is not completely negligible there.  The $n_e$ obtained from the hydrogen level population calculation is slightly larger in the majority of the regions except the part that is close to the inner boundary.  It is because that, in the hydrogen level population calculation, the secondary ionization due to metal species (see Equation~\ref{eq:2nd}) is not included, and the recombinations are only included up to $n=6$, while the case B recombination rate used in the hydrostatic model sums over all levels.

The number density of H($n=3$) obtained from the level population is also shown in the Figure~\ref{fig:pop}.  Because an optically thin stellar \Ha\ intensity is applied and Ly$\beta$ radiation transfer is not carefully considered, the number density of H($n=3$) shown here is only a rough estimate.  Nevertheless, the low number density indicates that the Paschen series absorption features are unlikely to be observed.

\section{TRANSMISSION SPECTRUM}
\label{sec:transit}

For a planet at distance $d$ from the observer, with uniform intensity $I_\nu$ over the stellar disk, the measured flux is
\begin{equation}
F_\nu  =  \frac{2\pi I_\nu}{d^2} \int_0^{R_\star} e^{-\tau_\nu(b)} b \dd b ,
\end{equation}
where $\tau_\nu(b)$ is the optical depth along a trajectory associated with the impact parameter $b$.  The optical depth can be divided into a continuum part, $\tau_c(b)$, which is independent of frequency over the line, and the line opacity part due to absorption by H(2$\ell$), 
\begin{equation}
\tau_{l,\nu}(b)=2\int_0^{\sqrt{R_{\rm top}^2-b^2}}(n_{\rm 2s}\sigma_{\rm 2s}+n_{\rm 2p}\sigma_{\rm 2p})\dd s,
\end{equation}
where $s$ is the line of sight distance.
The continuum absorption is then approximated as complete for $b<R_p$ and zero for $b>R_p$. The continuum integral then becomes as
\begin{align}
F^{(c)}_{\nu}  &\equiv F_\nu-F_\nu^{(c)} \nonumber \\
&=   \frac{2\pi I_\nu}{d^2} \int_0^{R_\star} e^{-\tau_c} b \dd b  = I_{\nu} \frac{\pi (R_\star^2-R_p^2)}{d^2},
\end{align}
where $R_p$ is the radius of the planet due to the continuum opacity. The difference in flux due to total opacity and continuum opacity is then
\begin{equation}
\Delta F_\nu  =  \frac{2\pi I_\nu}{d^2} \int_0^{R_\star}  \left( e^{-\tau_c(b)-\tau_{l,\nu}(b)}  - e^{-\tau_c(b)} \right) b \dd b.
\end{equation}
The contribution from both terms is zero for $b<R_p$ due to the continuum opacity, and there is no continuum absorption outside that range, and so this expression can be rewritten
\begin{equation}
  \label{eq:10}
\Delta F_\nu  =  \frac{2\pi I_\nu}{d^2} \int_{R_p}^{R_\star}  \left( e^{-\tau_{l,\nu}(b)}  - 1 \right) b \dd b.
\end{equation}
Equivalent to the transmission spectrum defined in the observations (e.g. \citet{Cauley2015}), the fractional change in flux, relative to the continuum integral at the same frequency,  is then
\begin{align}
\frac{\Delta F}{F}(\nu)  &\equiv  \frac{\Delta F_\nu}{F^{(c)}_{\nu}} \nonumber \\
&= \frac{2}{R_\star^2-R_p^2} \int_{R_p}^{R_\star}  \left(e^{-\tau_{l,\nu}(b)}  - 1 \right) b \dd b.
\end{align}
The ratio $\Delta F/F$ will be referred to as the model transmission spectrum.

\subsection{{\rm \Ha} and {\rm \Hb} Transmission Spectrum}
\label{sec:ha}

\begin{figure}
\begin{center}
\includegraphics[width=0.5\textwidth]{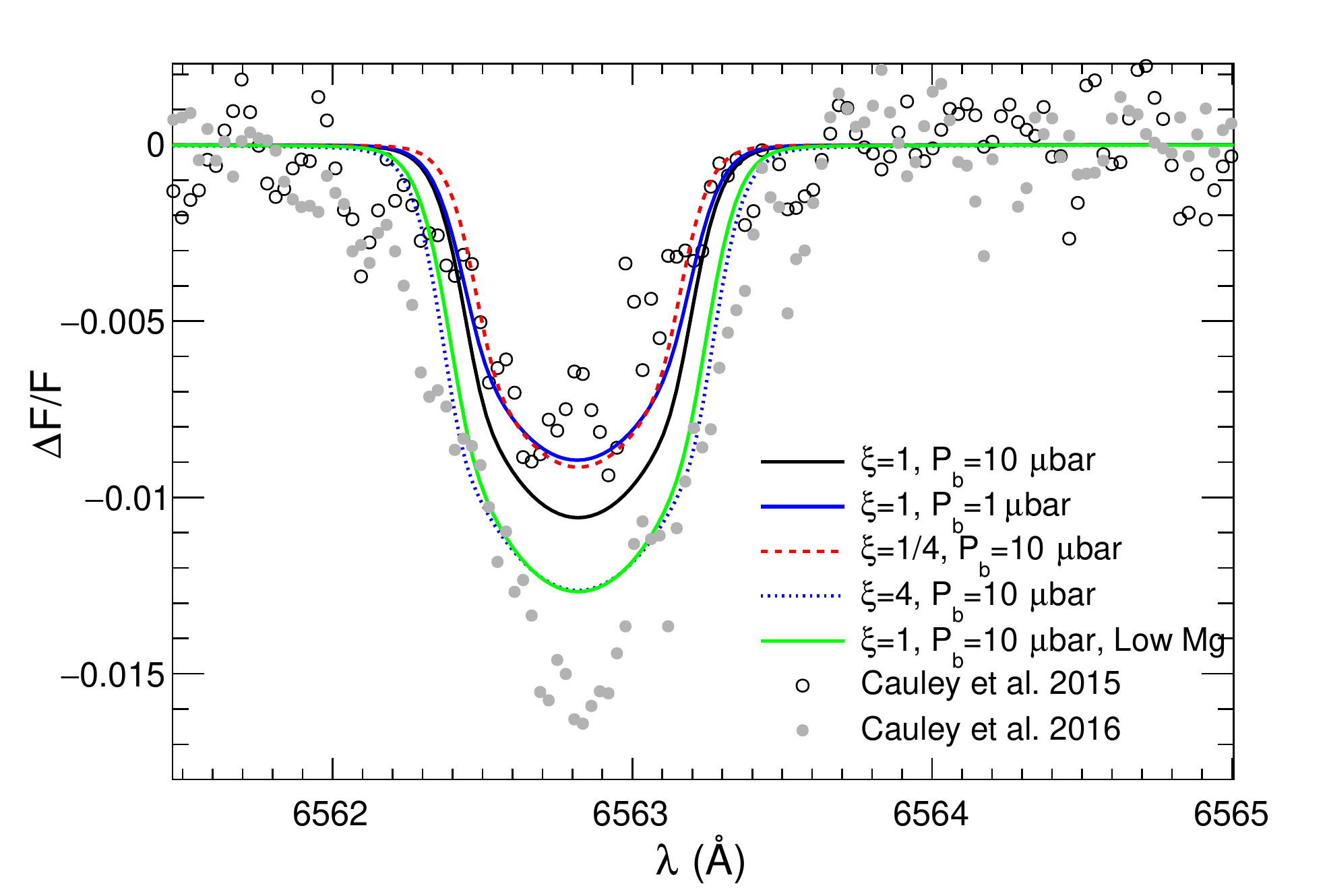}
\caption{\Ha\ transmission spectrum.  The black solid line shows the transmission spectrum of the fiducial model.  The blue solid line shows the model with atomic layer base pressure $P_b=1~\mu$bar.  Black and red dashed lines show the model with stellar LyC multiplier factor $\xi=1/4$ and $\xi=4$ respectively.  The red solid line shows the model without Mg.  Circles in the plot are observational data from \citet{Cauley2015, Cauley2016}.}\label{Halpha}
\end{center}
\end{figure}

\begin{figure}
\begin{center}
\includegraphics[width=0.5\textwidth]{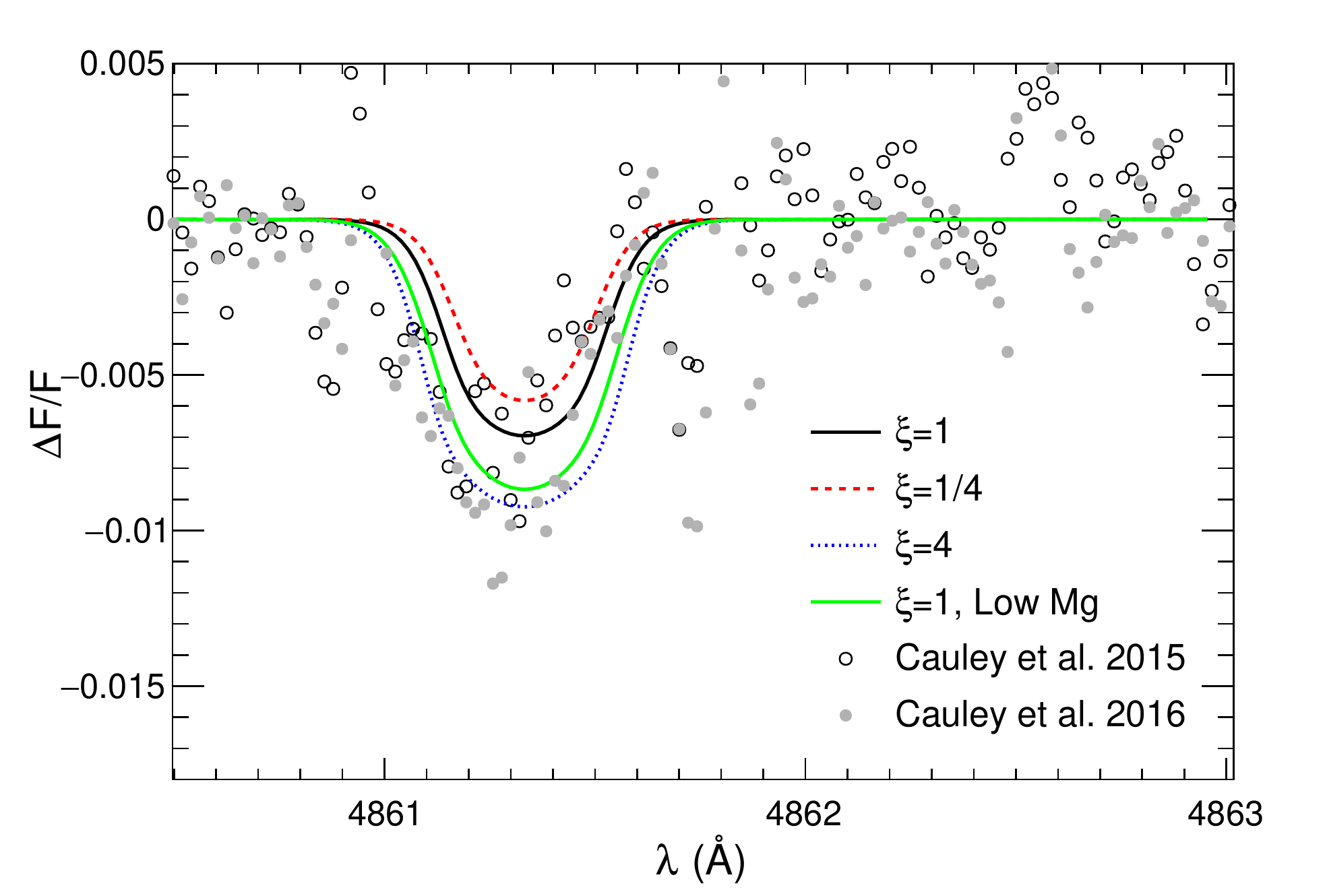}
\caption{Comparison of the \Hb\ transmission spectrum observed by \citet{Cauley2015} and \citet{Cauley2016}, to the calculated transmission spectrum of the fiducial model, as well as models with different LyC boost factor $\xi$ and Mg abundance.}\label{Hbeta}
\end{center}
\end{figure}

\begin{figure}
\begin{center}
\includegraphics[width=0.5\textwidth]{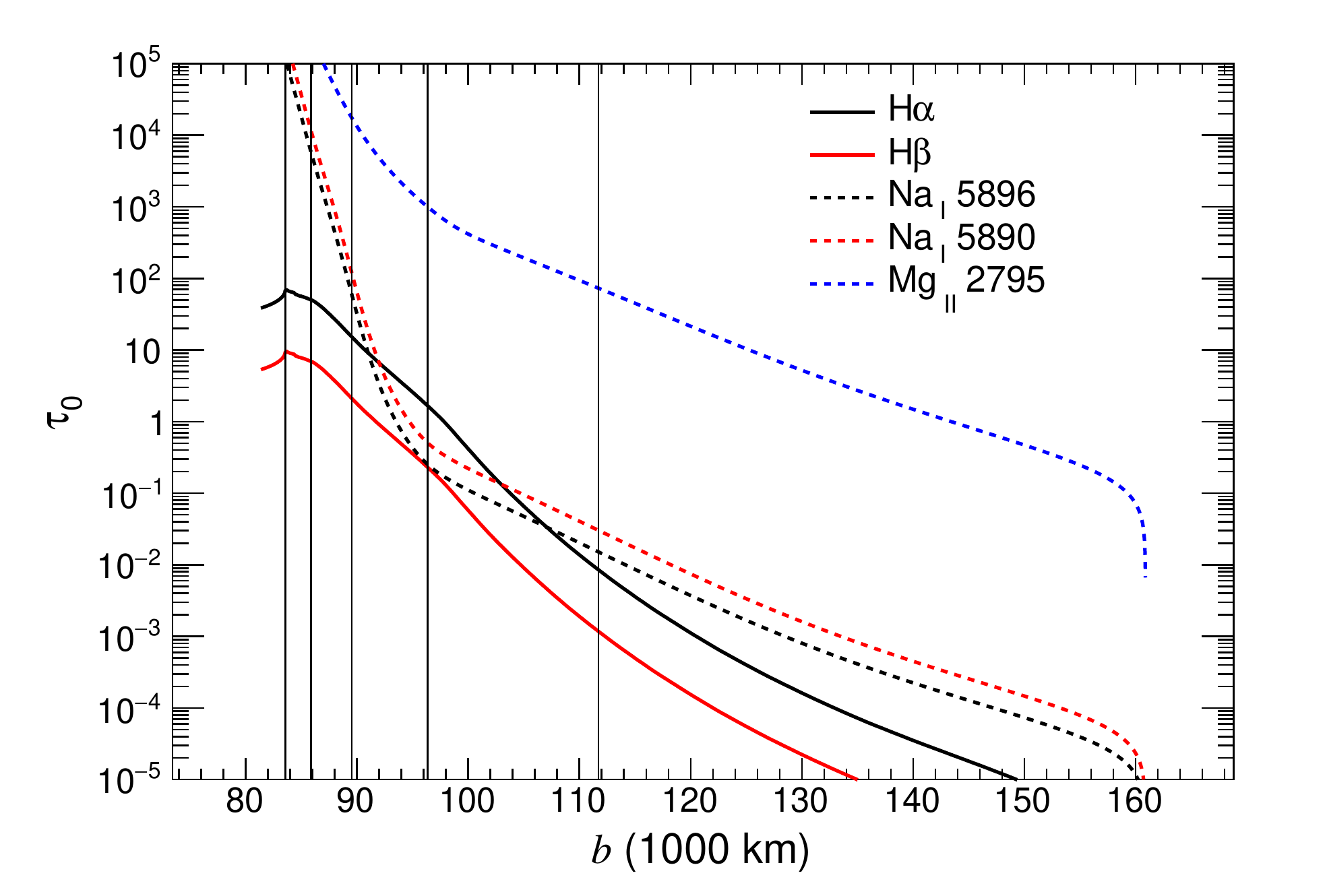}
\caption{\Ha, \Hb, Na D doublet, and \ion{Mg}{2} 2795 line center optical depth ($\tau_0$) versus impact parameter ($b$).  The vertical solid lines show the location of pressure levels $10~\mu$bar, $1~\mu$bar, $0.1~\mu$bar, $0.01~\mu$bar, and $0.001~\mu$bar from left to right.}\label{tau_b}
\end{center}
\end{figure}

\begin{figure}
\begin{center}
\includegraphics[width=0.5\textwidth]{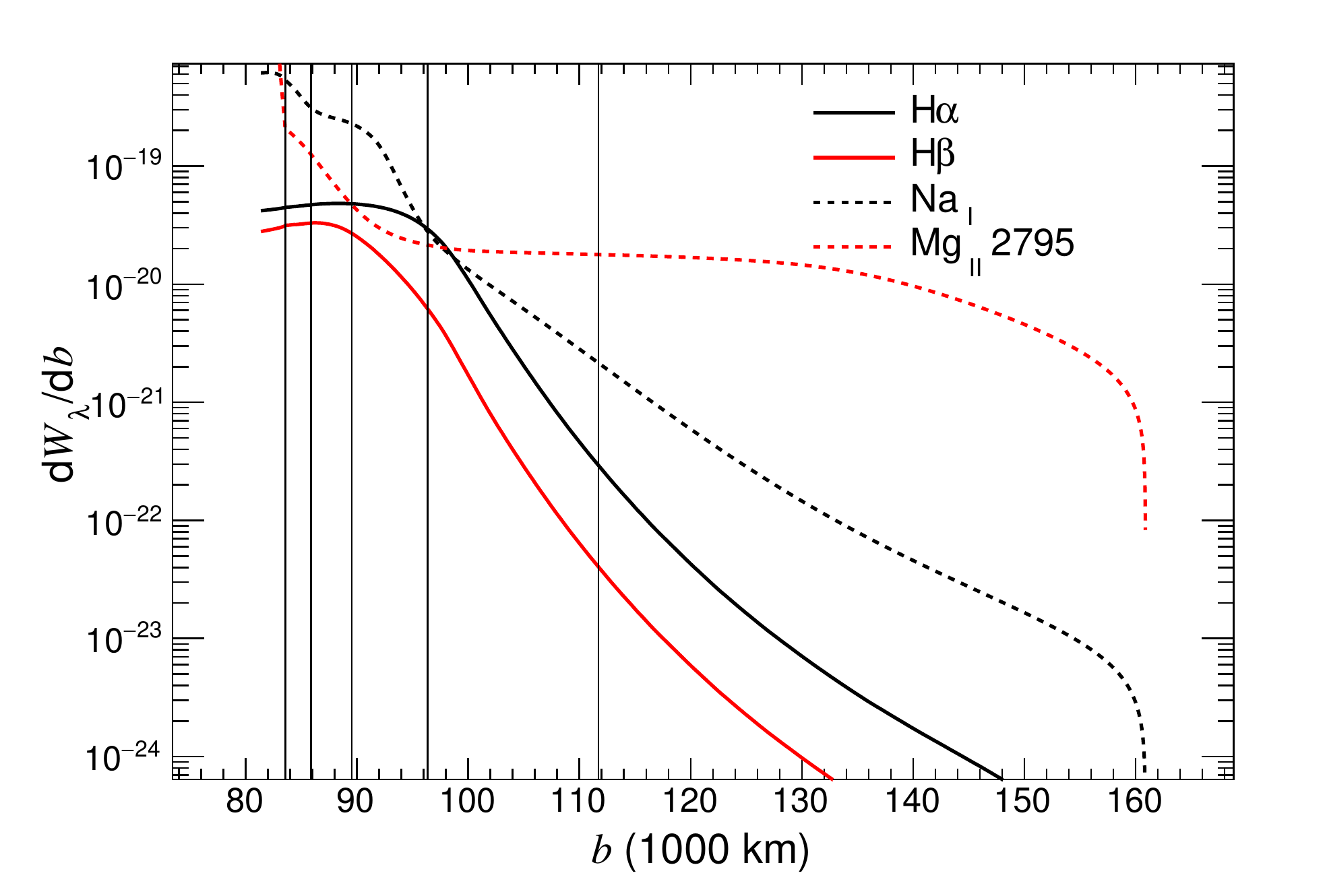}
\caption{\Ha\ and Na D transmission spectrum equivalent width per unit impact parameter ($\dd W_\lambda/\dd b$) (defined in Equation~\ref{eq:4}) versus b.
The contribution to $W_\lambda$ by Na D doublet are summed together.
The vertical solid lines show the location of $10~\mu$bar, $1~\mu$bar, $0.1~\mu$bar, $0.01~\mu$bar, and $0.001~\mu$bar from left to right.}\label{dWdb}
\end{center}
\end{figure}

From the discussion in Section~\ref{MC}, the \Lya\ intensity is small in the molecular layer at $r<R_b$. As a result, the H(2$\ell$) density there is small, and the region between $R_p$ and $R_b$ is transparent to \Ha.  

Figure~\ref{Halpha} shows the model \Ha\ transmission spectrum and the data from \citet{Cauley2015, Cauley2016}. 
We do not include the results of \citet{Jensen} since these observations were not performed across a single transit.
The fiducial model discussed in Section~\ref{sec:fiducial} is given by the black line labeled ``$\xi=1, P_{\rm b}=10\, \mu$bar". Given the noise in the data, the fiducial model is in broad agreement for both the line center absorption depth and the line width. 
The double-peak feature in \citet{Cauley2016}'s observations, whose amplitude is similar to the fluctuation in the continuum wavelength, cannot be explained by the model.
The wavelengths have been corrected for the index of refraction of air at ``standard condition'', $n_{\rm H\alpha}=1.0002762$, according to \citet{Refractive}. The plot also shows the effect of a different LyC flux, as denoted by lines with a different value of the factor $\xi$ and the metallicity, which will be discussed in the Section~\ref{sec:euv} and \ref{sec:z}, respectively. 

The base pressure, $P_{\rm b}$, is not self-consistently determined in this study. In order to investigate the dependence of the transmission spectrum on this parameter, the blue solid line labeled ``$\xi=1, P_{\rm b}=1\, \mu$bar" shows a model with the base of the atomic layer at $P_b=1~\rm \mu bar$. The line center transit depth is smaller by $\simeq 20\%$ for $P_b=1~\rm \mu bar$. Changing this boundary causes only small changes the atmosphere properties, so the \Ha\ becomes optically thick at approximately the same pressure.  
However, because the scale height between $1~\rm \mu bar$ and $10~\rm \mu bar$ significantly decreases after switching to lower temperature and larger mean molecular weight, the radius in the atomic layer that corresponds to the same pressure becomes smaller, which leads to a smaller transit depth.  Although we do not expect the transition from atomic layer to molecular layer to be as high up as $1~\rm \mu bar$ based on Figure~\ref{fig:n_P}, a more physical molecular model is required to produce a more precise transmission spectrum.  

Figure~\ref{Hbeta} shows the model \Hb\ transmission spectrum and the data from \citet{Cauley2015, Cauley2016}.  
While the model roughly agrees with the observation, there is an extra absorption on the blue side of the line, which cannot explained by the model.
Because of the smaller cross section, \Hb\ line probes a deeper region in the atmosphere compared to \Ha.
The lower temperature there leads to a narrower line width. 
Compared to \Ha, the \Hb\ observations have larger uncertainty and less significant transit depth variation.

Figure~\ref{tau_b} shows the line center optical depth of the fiducial model versus impact parameter $b$. The \Ha\ line center optical depth reaches the maximum value $\sim 70$ for $b=R_{\rm b}$, the base of the atomic layer.  Although the optical depth of \Ha\ slightly decreases inward, due to the (assumed) transparent molecular layer, the optical depth is still much larger than 1 all the way to the continuum radius $b=R_{\rm p}$.  \citet{Lecavelier08} showed that the optical depth at the effective radius is $\tau_{eq} \simeq 0.56$\footnote{With the approximation of the uniform mixing ratio and isothermal thin atmosphere, the integral in Equation~\ref{eq:10} may be expanded in a series as $\int_0^\infty \dd u \left( 1 - e^{-\beta e^{-u}} \right)  \simeq   \ln(\beta) + \gamma +  {\cal O} (\beta^{-2})$, where $\gamma \simeq 0.577$ is the Euler-Mascheroni constant and $\beta \simeq n(R_p)\sigma\sqrt{2\pi R_pH}$.  This formula is valid for $\beta \gg 1$. This expansion then gives $\tau_{\rm eq}=e^{-\gamma} \simeq 0.561$, in good agreement with \citet{Lecavelier08}.},
and is not sensitive to the details of the atmospheric structure.  In the fiducial model, the effective radius is $9.93\times 10^9~\rm cm$, corresponding to an \Ha\ optical depth $\tau=0.55$ and pressure $P=5.2\times 10^{-3}~\rm \mu bar$.  The optical depth drops to below $\sim 10^{-2}$ at a pressure $10^{-3}~\rm \mu bar$, which means the contribution to \Ha\ absorption from the atmosphere above this level is small.  
Determined by the ratio of oscillator strength and wavelength, the ratio between the optical depth of \Ha\ and \Hb\ is 7.3.

To indicate the vertical distribution of \Ha\ and Na D absorption by the atmosphere, Figure~\ref{dWdb} shows the equivalent width contributed by an annulus of atmosphere with radius $b$, defined as
\begin{equation}
  \label{eq:4}
  \frac{\dd W_\lambda}{\dd b}=\frac{2b}{R_\star^2-R_p^2}\int \left(1-e^{-\tau_{l,\nu}(b)}\right)\dd \lambda.
\end{equation}
For impact parameters in the range $R_{\rm p} < b < R_{\rm b}$, which go through the molecular layer, the \Ha\ line center optical depth is $\simeq 70$, and the absorption in the Lorentzian damping wing is negligible.  Therefore, the base of the atomic layer is in the flat portion of the curve of growth~\citep{Draine}.  The contribution to the equivalent width decreases slowly inward in this part of the atmosphere because of the smaller annulus radius and the lower temperature.
It is shown that the atomic layer of the atmosphere has an approximately uniform contribution to \Ha\ absorption, while the absorption of \Hb\ is dominated by the region of $P\gtrsim 0.1~\rm \mu bar$.
In principle, high quality \Ha\ and \Hb\ observations can be a good tracer for the vertical structure of the atomic layer.

\subsection{{\rm Na D} Transmission Spectrum}

\begin{figure}
\begin{center}
\includegraphics[width=0.5\textwidth]{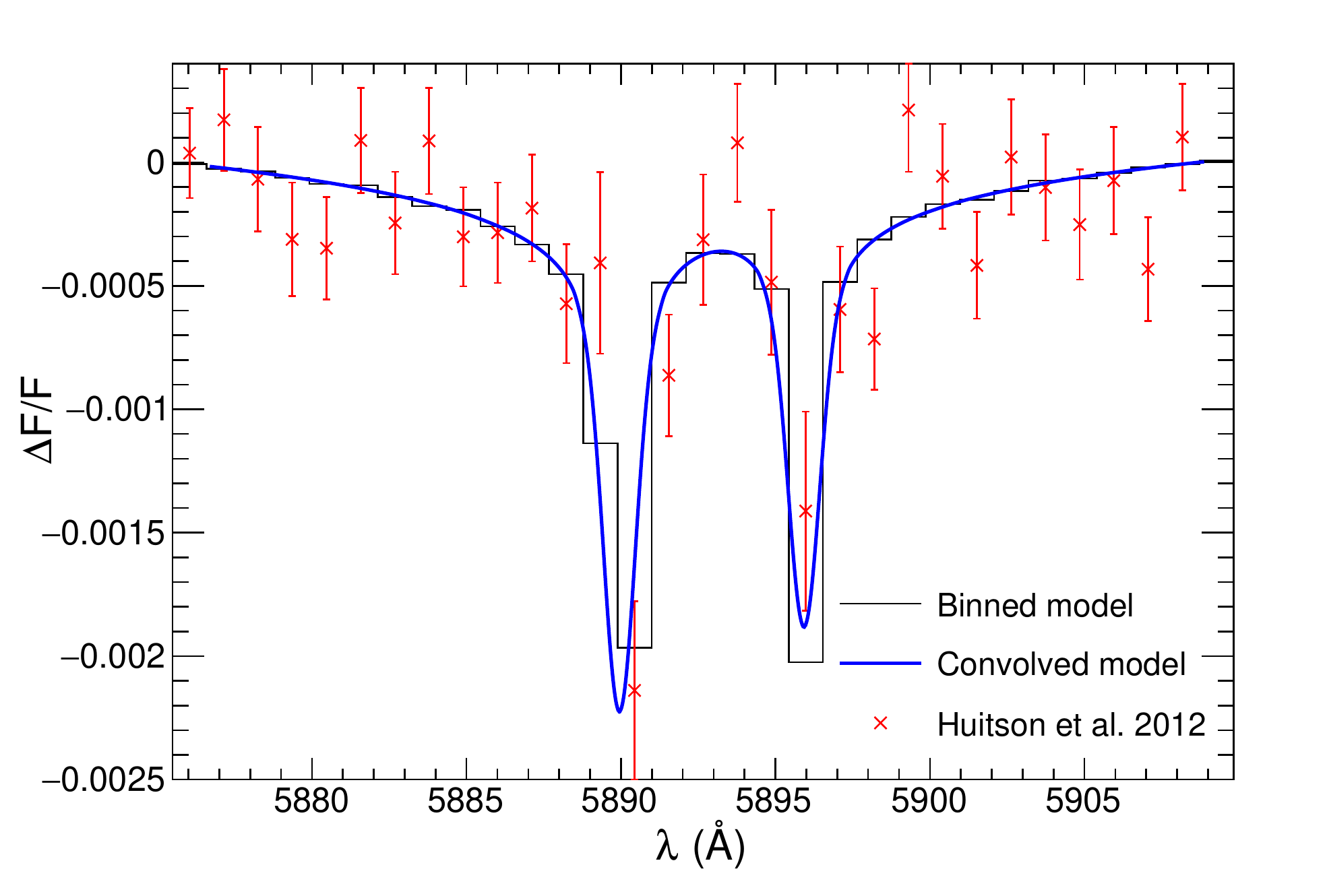}
\caption{Comparison of the Na D doublet transmission spectrum observed by \citet{Huitson} to the transmission spectrum of the fiducial model ($\xi=1, P_b=10~\mu$bar).  The histogram shown as a black line is the modeled spectrum binned to the instrument resolution.  The blue curve shows the result of the modeled spectrum convolved with a Gaussian profile with FWHM matching the instrument resolution.}\label{Huitson}
\end{center}
\end{figure}

\begin{figure}
\begin{center}
\includegraphics[width=0.5\textwidth]{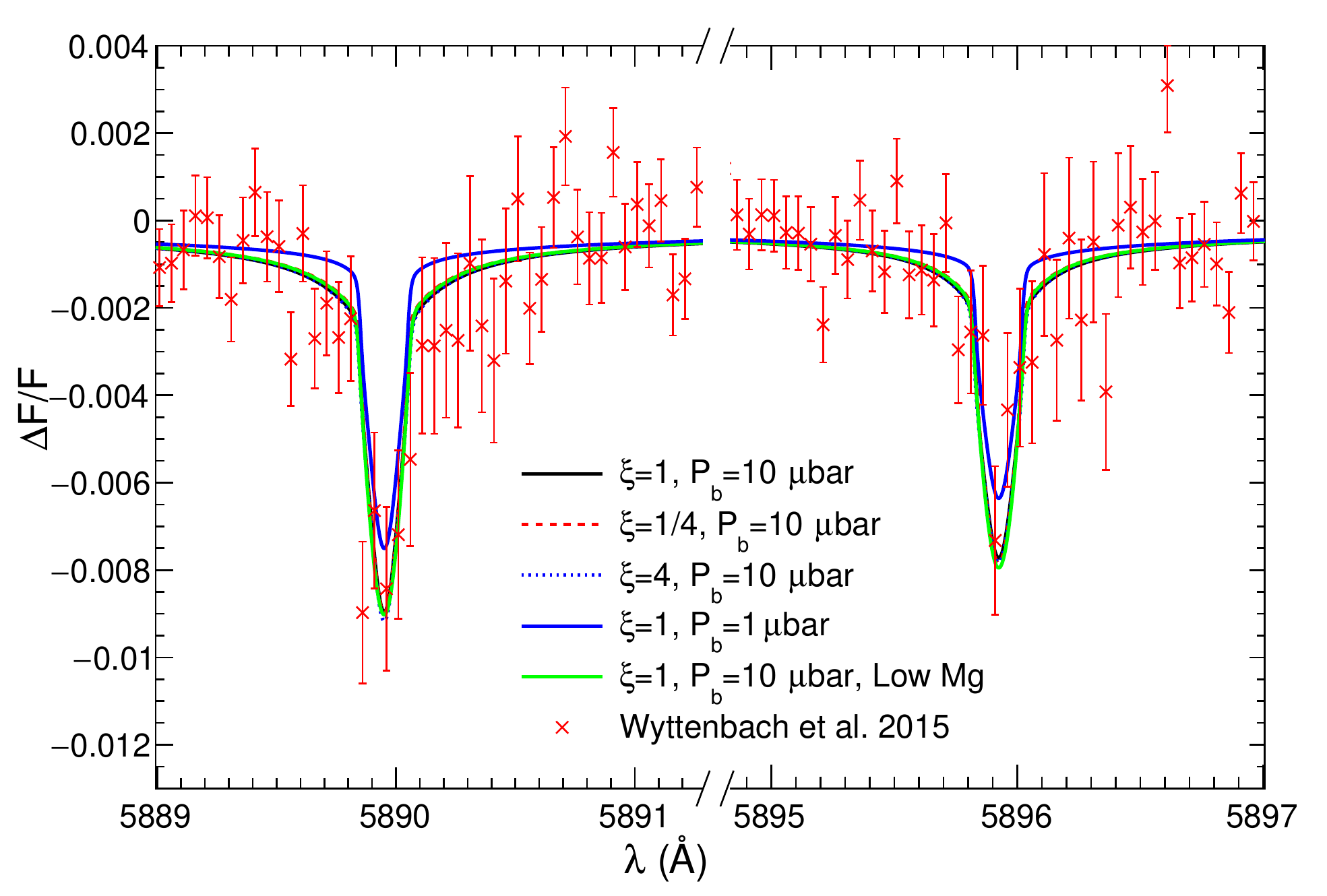}
\caption{Comparison of the Na D doublet transmission spectrum observed by \citet{Wyttenbach} binned by $5\times$, to the calculated transmission spectrum of the fiducial model, as well as models with different atomic layer base pressures $P_b$, LyC boost factor $\xi$ and metallicity.  Note the break in the x-axis.  The spectra of the three models with $P_b=10\rm~\mu bar$ and solar abundance are nearly overlap on each other.}
\label{Wyttenbach}
\end{center}
\end{figure}

Unlike H$\alpha$, the Na D doublet lines are absorbed by ground state Na, which has high density deep in the atmosphere. The molecular layer is not treated in detail here, rather a simple model with constant (solar abundance) mixing ratio and temperature $T=1140$ K is used.  Because Na and Mg are extremely optically thick in the molecular layer, their number density is not important in determining the line profile.
For simplicity, number density of Na and Mg are assumed to be equal to the value at the very base layer of the model.  Similar to \Lya, Na D photons undergo a resonant scattering process in the atmosphere, at least at pressures sufficiently low that collisional de-excitation and collisional broadening are negligible.  An accurate model of the Na D transmission spectrum requires treatment of resonant scattering by Na\UI, as well as true absorption and emission by the atmosphere, which is beyond the scope of this paper.  Instead, as was done for \Ha\ in this work, a simple $e^{-\tau_{l,\nu}(b)}$ absorption will be used to compute the transmission spectrum.  Because the line wings of the Na D doublet are overlapping, $\tau_{l,\nu}(b)$ uses the sum of the cross sections for each line of the doublet, evaluated at frequency $\nu$. 

Figure~\ref{tau_b} shows that the optical depths of the Na D doublet lines reach $\tau \sim 0.5$ at $P\sim 10^{-2}~\rm \mu bar$, comparable to that of \Ha, agreeing with the inference made previously based on the similar transit depth of \Ha\ and Na D.
At this altitude, the temperature is $\sim 8500$ K, comparable to the analytic estimate using the difference between the Na D doublet transit depths discussed in Section~\ref{sec:analytic}.  
The optical depth of the Na D doublet become much larger than unity below $1~\mu$bar.  Thus Na D absorption by the atmosphere near the base of the atomic layer is in the damped portion of the curve of growth, which explains the large contribution to the equivalent width shown in Figure~\ref{dWdb}.  
However, because of the slow transit depth variation with frequency on the damping wing, it is difficult to distinguish the Na D damping wings from possible additional sources of continuum opacity or observational error bars.  The presence of clouds or hazes would further complicate the detection of this lower portion of the atmosphere, in spite of its large equivalent width contribution.  
The portion of the curve that is deeper than $10~\mu$bar has no practical meaning because the value is limited by the wavelength integration range of the equivalent width.

Figure~\ref{Huitson} compares the observed Na D transmission spectrum from \citet{Huitson} and the fiducial model. The wavelength has been corrected for the index of refraction of air, $n_{\rm Na D}=1.0002771$ according to \citet{Refractive}.  The spectral resolution of the Space Telescope Imaging Spectrograph (STIS) G750M grating aboard the HST used in this observation is $\sim 6$ times broader than the full width at half maximum (FWHM) of each line in the Na D doublet.  Two methods are used to compare the model spectrum with the low spectral resolution observation.  The first method is described in \citet{Huitson}, and is shown as a black histogram. The model spectrum is binned to the STIS instrument resolution, 2 pixels, since the G750M grating gives a resolution $\sim$2 pixels at 5893~\AA.  Care is required since the absorption depth of the binned spectrum near each line center depends on the wavelength range used for binning.  In view of this, the second method, shown as a blue solid curve, convolves the model spectrum with a Gaussian profile with FWHM matching the instrument resolution (2 pixel widths).  For both methods, to imitate the process of normalization to the continuum outside the regions of interest which been done in observational data reducing, the Na absorption depth at 5912~\AA\ is subtracted out and treated as the continuum.

Figure~\ref{Wyttenbach} compares the high resolution Na D transmission spectrum observed by \citet{Wyttenbach} with the fiducial model, as well as models with different atomic layer base pressures $P_b$, $\xi$ and metallicity, equivalent to the models shown in Figure~\ref{Halpha}.  In comparing to the \Ha\ transmission spectrum, recall that the line width of Na is narrower as compared to \Ha\ due to the larger mean atomic weight of Na.  
To reduce the noise, the data plotted are binned by 5$\times$.  The resulting $0.05 \AArm$ bin width is equal to the FWHM of the average spectrograph line spread function, and is $\sim 3$ times narrower than the Na D FWHM.
On top of the $2.3 \kms$ shift to the red which accounts for the systemic velocity, the data were shifted by $10 \kms$ to the red to cancel the observed blueshift from an unknown source described in \citet{Wyttenbach}.  Similar to the treatment in Figure~\ref{Huitson}, the correction from the index of refraction and continuum flux are made to the simulated spectrum.  No binning or convolving is required because the spectral features are well resolved.

The line center absorption depths generated by the models agree with the Na D spectrum in both observations roughly to the level of the observational error bars.  Note that there is a strong absorption feature on the red side of the line center which cannot be explained by the model.

\subsection{Retrieval of the Temperature Profile from {\rm Na D} Transmission Spectra}
\label{sec:retr-temp}

To measure the temperature versus altitude profile from the Na D transmission spectrum, $r(\lambda)$, \citet{Lecavelier08} applied the analytic model  
\begin{equation}
  \label{eq:5}
  r(\lambda)=H\ln(\sigma(\lambda))+\text{constant},
\end{equation}
which is derived for a plane-parallel isothermal atmosphere with uniform mixing ratio of Na\UI\ and scale height $H=k_{\rm B}T/\mu m_p g$.
Here $\sigma(\lambda)$ is the summed cross section from each line of the Na D doublet, and a Voigt profile at the local temperature is used. The wavelength-independent constant term is determined by the radius at continuum wavelengths. 
If the temperature and abundance vary slowly with altitude, an approximate scale height $H$ of the atmosphere at a certain radius can be derived from $H=(\dd r/\dd \lambda)/(\dd \ln \sigma/\dd \lambda)$.
Then, applying a mean molecular weight $\mu$, the local temperature $T(r)$ at $r(\lambda)$ can be computed from the fitted value of $H$.

Methods similar to this were applied by \citet{Huitson} and \citet{Wyttenbach} to measure the upper atmosphere temperature from their observed Na transmission spectra.
The atmosphere was assumed to be molecular with $\mu=2.3$ in both studies.
To decrease the uncertainty of the temperature measurement due to noise in the observed $r(\lambda)$ profile, \citet{Huitson} broke the spectrum into small wavelength intervals, and fit $r(\lambda)$ in each interval by varying $H$ and the constant.
\citet{Wyttenbach} also broke the spectrum into intervals, and fit for $H$ in each spectral region, but with a fixed value for the constant term in Equation~\ref{eq:5} in each interval.
As a result, if connecting the fitting curves from separate wavelength ranges together, the joined curve is not continuous in both the slope and value of the transit depth at the boundary between adjacent wavelength ranges.

The measured temperatures near line center in both studies are more than a factor of 2 lower than the model temperature here (see Figure~\ref{Retrieve}) over the relevant region of the atmosphere.
This is in spite of their using a mean molecular weight $\mu=2.3$, which assumes molecular hydrogen, while here the mean molecular weight in the atomic layer is closer to $\mu \simeq 1.3$, smaller by a factor of 2.
Such a large difference in temperature cannot be explained by the abundance variations due to ionization seen in Figure \ref{fig:n_P}.

\begin{figure}
\begin{center}
\includegraphics[width=0.5\textwidth]{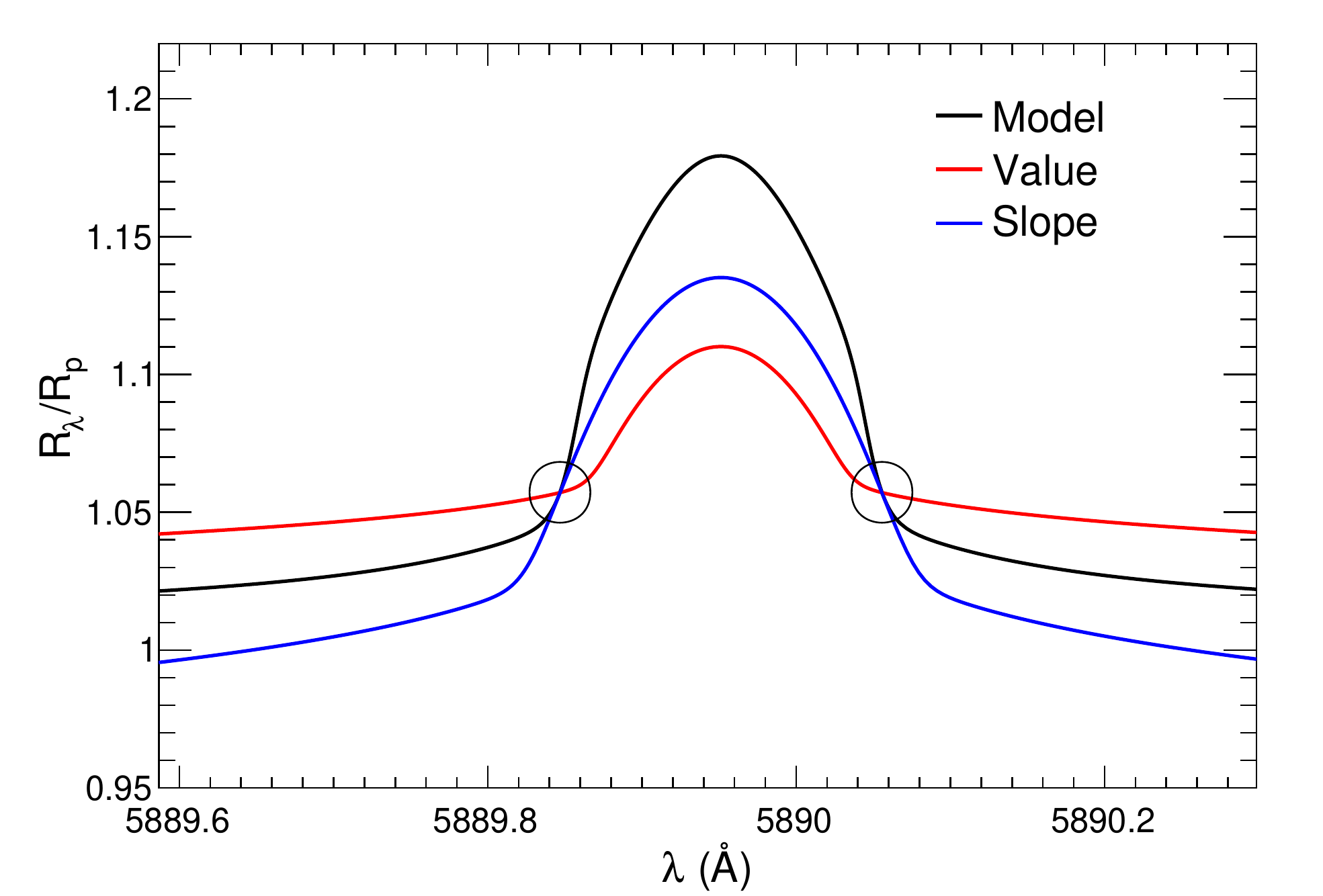}
\caption{Comparison of the fiducial model transmission spectrum and two different fits at a single point (circled) using the isothermal transit radius approximation in Equation \ref{eq:5}.  The black line shows the transit radius of the Na\UI\ 5890 line for the fiducial ``$\xi=1, P_{\rm b}=10\mu\, \rm bar$" model (see Figures~\ref{Huitson} and \ref{Wyttenbach}).  The red curve is chosen to agree with the value of the transit radius at $r(\lambda)=1.057R_p$, and with the continuum $r(\lambda)$ on the line wing.  This requires a temperature $T=2830$ K, assuming $\mu=2.3$.  The blue curve is chosen to have the same slope as the black curve at the circled point, which requires $T=6010$ K for $\mu=2.3$. }
\label{Fitting}
\end{center}
\end{figure}

\begin{figure}
\begin{center}
\includegraphics[width=0.5\textwidth]{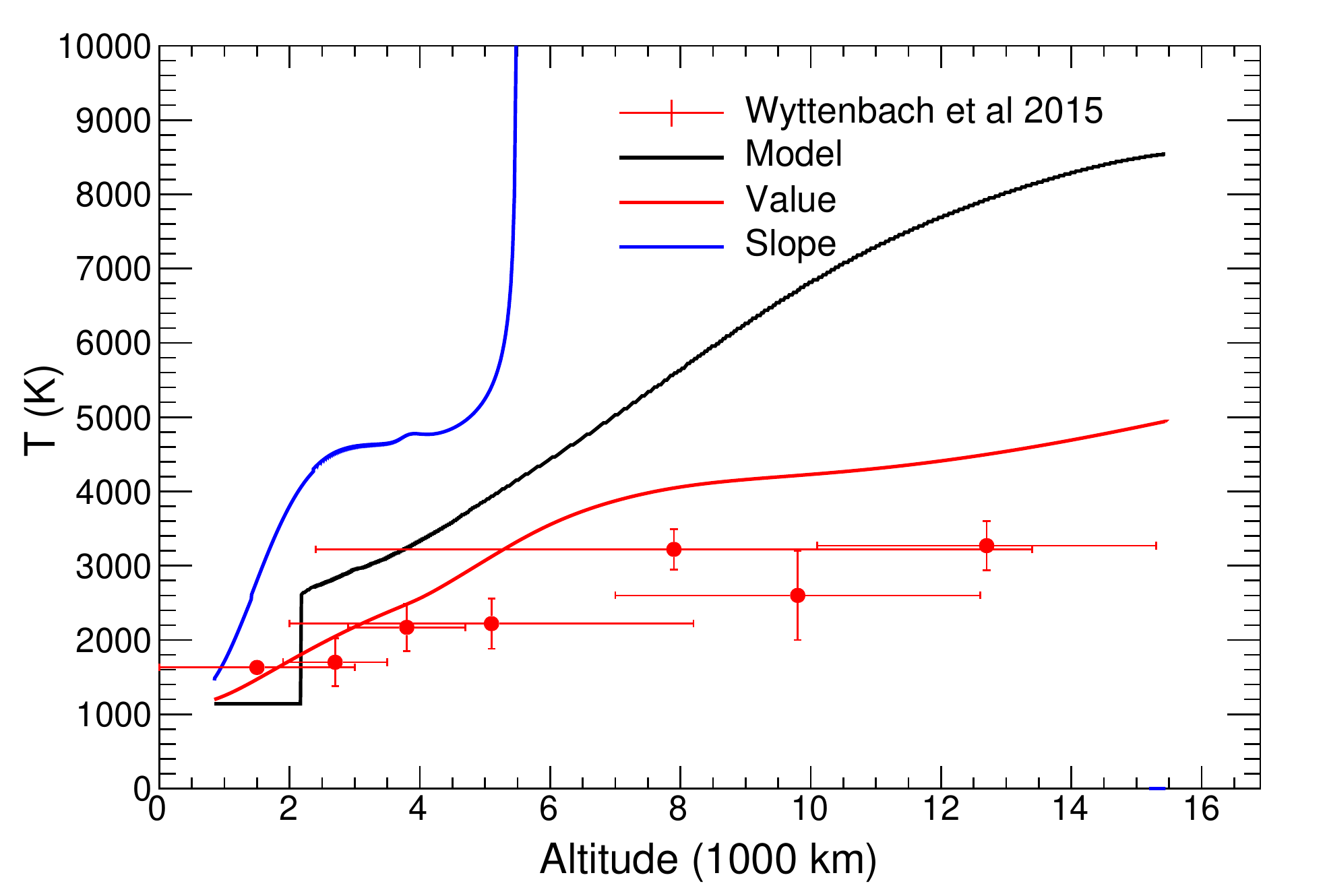}
\caption{Retrieved temperature profile using two different fitting strategies shown in Figure~\ref{Fitting}.  Temperature profile fitted from the observed Na D transmission spectrum obtained by \citet{Wyttenbach} is shown with red circles. }
\label{Retrieve}
\end{center}
\end{figure}

One indication that higher temperatures than found in \citet{Huitson} and \citet{Wyttenbach} are required comes from the inferred range of density between line center and line wing.  A lower temperature means a larger density difference between the base of the atmosphere, where the continuum forms, and higher altitudes where the line center forms.  An underestimate of this density decrease can be found by using the highest fitted temperature used in \citet{Wyttenbach}, $T=3270$ K, and mean molecular weight $\mu=2.3$.  Assuming the pressure is large, 1 bar at the continuum altitude, also errs on the side of high density higher in the atmosphere.  With these two assumptions, an isothermal atmosphere gives the pressure $5\times 10^{-5}~\rm \mu bar$ at the line center altitude $1.27\times 10^4$ km. 
To be optically thick to the Na\UI\ 5890 line, the ground state density must be $n_{\rm Na_I} \ga (\sigma_{\rm Na D2}\sqrt{2\pi r H})^{-1} \simeq 10^2 \rm~cm^{-3}$, where $\sigma_{\rm Na D2}$ is the line center cross section of  Na\UI\ 5890.  
This requires a $\sim 10^{-6}$ mixing ratio of Na\UI\, which means Na has to be mostly neutral at this altitude if the atmosphere is in solar abundance.  
However, because the atmosphere is optically thin to the stellar flux at this pressure, Na is significantly ionized (see the left hand side of Figure \ref{fig:n_P}).
Therefore, the highest temperature measured in \citet{Wyttenbach} may underestimate the line center temperature formed high in the atmosphere.

For sufficiently high spectral resolution data, $r(\lambda)$, with high signal to noise for each data point, and for an atmosphere which is nearly isothermal and with small abundance gradients, the temperature of the atmosphere will be accurately recovered using Equation~\ref{eq:5} in the plane-parallel limit.
However, in an atmosphere where temperature increases upward rapidly, this method tends to underestimate the temperature \citep{Wyttenbach}. The problem is exacerbated when the opacity is provided by the line's Doppler core at the altitudes of interest.

To better understand the retrieval of a temperature profile for the non-isothermal, non-constant abundance case, an example is given here to fit the fiducial model $r(\lambda)$ (see Figures~\ref{Huitson} and \ref{Wyttenbach}) with the isothermal profile in Equation~\ref{eq:5}. This eliminates measurement errors in the data, and a fine enough grid of points is used so that numerical error is negligible. The black curve in Figure~\ref{Fitting} shows the fiducial model for the Na\UI\ 5890 absorption profile, the same as in Figure~\ref{Wyttenbach}. Two methods are used to fit Equation~\ref{eq:5} to the fiducial model.
``Method 1" is equivalent to that in \citet{Wyttenbach}.  The constant term in Equation~\ref{eq:5} is chosen in order that the absorption depth is 0 at 5912~\AA, on the line wing.
The temperature $T(r)$ at each radius $r(\lambda)$ is determined by matching the value of the transit radius using Equation~\ref{eq:5} to the fiducial model.
A mean molecular weight $\mu=2.3$ is used, as in \citet{Wyttenbach}.
Figure~\ref{Fitting} shows an example of a Method 1 fit with $T=2830$ K, which matches the value of the absorption depth at $R=1.057R_p$. When the value is fitted, the slope will be smaller than that of the fiducial model.
``Method 2" is equivalent to that in \citet{Huitson}.
By adjusting $H$ and the constant term at each $r(\lambda)$, Equation~\ref{eq:5} is used to match the slope of the fiducial model transit radius.
Again $\mu=2.3$ is used.
Figure~\ref{Fitting} shows an example of a Method 2 fit with $T=6010$ that is tangent to the fiducial model $r(\lambda)$ curve at $R=1.057R_p$.

The retrieved temperature profiles for Method 1 and Method 2 are compared to the fiducial model temperature profile in Figure~\ref{Retrieve}.
For comparison, the Method 1 temperature profile estimated from the data by \citet{Wyttenbach} is shown as the points with error bars. Given that there is no numerical noise in this example, as the isothermal $r(\lambda)$ is fit to a theoretical model, the disagreement between the Method 1 and Method 2 fits and the true temperature profile is quite large. The disagreement would be even larger if the more appropriate $\mu \simeq 1.3$ was used near line center.
Method 1 can reasonably retrieve the temperature in the molecular layer, where the fiducial model temperature is constant and $\mu=2.3$.
However, the retrieved temperature in the atomic layer is lower than the fiducial model where the temperature increases outward.
Although the retrieved temperature is still higher than the points from \citet{Wyttenbach}, this example partially explains the lower inferred temperature in that work as compared to the fiducial model in this work.

Retrieving the temperature from a high resolution spectrum using Method 2 will significantly overestimate the temperature.  In the Doppler core, Equation \ref{eq:5} gives
\begin{eqnarray}
r(\lambda) - r(\lambda_0) & =& - H \left( \frac{\Delta \nu}{\Delta \nu_D} \right)^2  \nonumber \\
&\simeq &- 6000\, {\rm km} \left( \frac{\Delta \lambda}{0.1\AArm}
  \right)^2 ,
\label{eq:isolinecenter}
\end{eqnarray}
independent of temperature.
Equation \ref{eq:isolinecenter} shows that the slope gets steeper further from line center.
This continues until the damping wing is reached, where the slope becomes more shallow. 
Therefore, Equation~\ref{eq:isolinecenter}, evaluated near the core-wing boundary, gives the maximum slope of transit radius with $\Delta \lambda$ for the isothermal profile.
By contrast, the fiducial model $r(\lambda)$ is steeper than Equation \ref{eq:isolinecenter} in the line core because the absorption by a higher and hotter atmosphere layer produces a broader absorption than if the temperature is constant.
As the result, the slope of a section of the fiducial model near the line core is too steep and Equation \ref{eq:5} cannot produce such a steep slope for any temperature.
In addition, near the line core, the slope of $r(\lambda)$ for the fiducial model also depends on the Na\UI\ abundance gradient.
Collisional ionization by thermal electrons and \Lya\ photoionization decrease the Na ionization fraction at level above $0.1 \rm~\mu bar$ (see Figure \ref{fig:n_P}), which can decrease the slope near the line center.

These results suggest that the isothermal model may not accurately retrieve a rapidly rising temperature profile (see also \citet{Heng}). In comparison, fitting the whole wavelength range with an atmosphere model contains several isothermal layers, or a single layer with a continuous temperature profile may better constrain the atmosphere temperature.

\subsection{{\rm Mg} Transmission Spectrum}

\begin{figure}
\begin{center}
\includegraphics[width=0.5\textwidth]{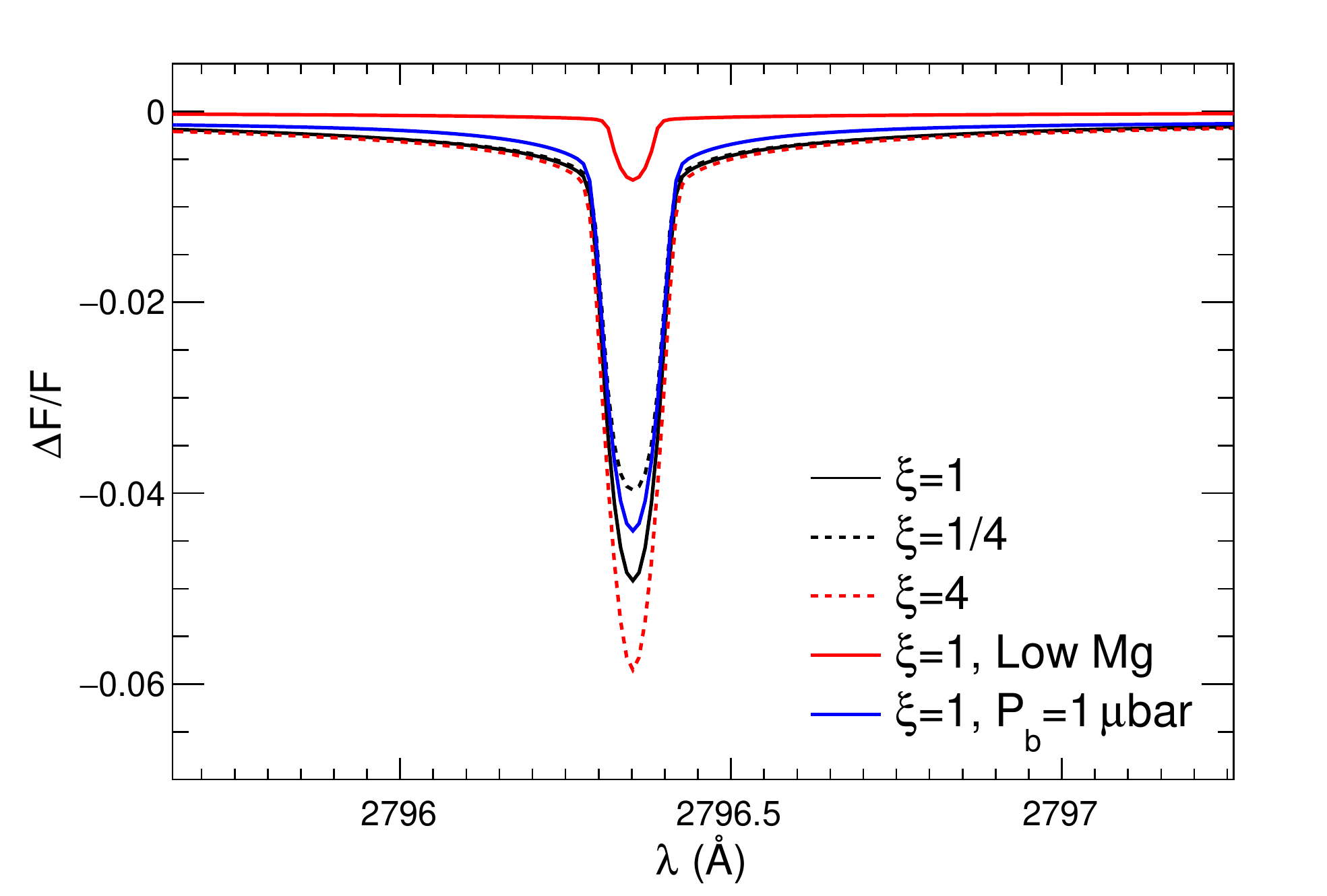}
\caption{The calculated \ion{Mg}{2} 2795 transmission spectrum of the fiducial model, as well as models with different LyC boost factor $\xi$ and Mg abundance.}\label{MgII}
\end{center}
\end{figure}

Figure~\ref{MgII} shows the model predicted transmission spectrum of \ion{Mg}{2} 2795, one of the \ion{Mg}{2} doublets.  No correction of index of refraction has been applied.
Assuming solar metalicity, almost the entire atmosphere in the simulation region is optically thick to Mg lines at the line center, because of the high abundance and shallower dependence on pressure (see Figure~\ref{fig:n_P}).
In this case, the line center transit depth is $\sim 5\%$, 3 times larger than the planet transit depth in continuum.
In contrast, assuming the abundance of Mg is $10^{-4}$ of the solar value, the transit depth is $\sim 0.8\%$. 
The model predicts very similar transmission spectra for \ion{Mg}{2} 2803 and \ion{Mg}{1} 2852 line.
Because of the probable large transit depth and its strong impact on the physical properties of the atmosphere, the transmission spectra of Mg resonant lines in middle UV might be a good target to constrain the hot Jupiter upper atmosphere.

\subsection{Impact of LyC Flux on Transit Depth}
\label{sec:euv}

HD 189733 is known to be an active star \citep{Boisse, Pillitteri2014, Pillitteri2015}.  To show the dependence of the transmission spectrum on the stellar EUV/X-ray flux, two more models with an extra LyC flux multiplier factor $\xi=1/4$ and $\xi=4$ are applied. 
Figure~\ref{Halpha} and \ref{Hbeta} show that a stronger LyC flux will make the \Ha\ and \Hb\ transit depth deeper.
This is consistent with the conclusion in \citet{Cauley2017} that the \Ha\ transit showing the largest absorption value occurs when the star is the most active.
In comparison, Figure~\ref{Wyttenbach} shows that LyC flux has no effect on the NaD transmission spectrum.
The reason of this difference is that the Na D transmission spectrum depends on the temperature or scale height of the atmosphere as well as Na ionization fraction below $10^{-2}~\rm \mu bar$.
Above the level of $10^{-1}~\rm \mu bar$, collisional ionization dominate Na ionization.
Although the temperature of the atmosphere increases with the $\xi$, the effect of increasing scale height is canceled by the Na higher ionization fraction.  
In contrast, the ionization of H is not sensitive to the temperature.
Instead, Balmer lines depend on the \Lya\ intensity in this region, which is larger for a strong LyC flux environment.
The Balmer lines transmission spectra also become slightly broader in the strong LyC flux case, because the lines become optically thick at higher and hotter part of the atmosphere in this case.
This indicates that Balmer lines are the better tracer of atmosphere temperature compared to Na doublets.
Comparing the variability of the transit depths of the \Ha\ and Na D lines is a possible method to break the degeneracy between the transit depth variability due to blocking an active region on the star surface and the change in the atmosphere due to stellar activity.

\subsection{Impact of Metallicity on Transit Depth}
\label{sec:z}

The metallicity is crucial in the model presented in this paper, but its value is uncertain.  Since Mg is the dominant coolant in the model, a model that reduces the Mg abundance to $10^{-4}$ of the solar value is calculated to assess the effect of metallicity on the transmission spectrum.  Because the atmosphere is warmer and more extended without Mg cooling, the transit depths of \Ha\ and \Hb\ become deeper, as shown in Figures~\ref{Halpha} and \ref{Hbeta}.  The transit depth of Na is insensitive to the Mg abundance because of the trade off between atmosphere scale height and ionization fraction discussed in~\ref{sec:euv}.  The relatively large transit depth difference between two models with and without Mg indicates that high precision Balmer and Mg transmission spectrum measurements can constrain the metallicity in the upper atmosphere.  

\section{DISCUSSION}\label{discussion}

\subsection{Other Possible Cooling Mechanisms}
\label{sec:ad_cool}

Adiabatic cooling is another potential cooling mechanism discussed in the literature.  \citet{KoskinenI} constructed an atmosphere model for a similar hot Jupiter HD 209458b.  In their model, the stellar heating is mainly balanced by adiabatic cooling.  Compared with \HD\ system, the LyC flux of HD 209458 is weaker and the orbit of HD 209458b is further away from the star.  They also introduced a factor of 1/4 reduction on stellar flux to account for uniform day-night heat redistribution.  As a result of these differences, the heating rate of \citet{KoskinenI} HD 209458b atmosphere model is more than $20$ times smaller than the rate in our \HD\ model.  On the other hand, the adiabatic cooling does not differ much in two systems because of the similar mass loss rate~\citep{Murray-Clay}.  Therefore, the adiabatic cooling is unlikely to be the answer in the case of HD 189733b.

The effect of the adiabatic cooling of the model here can be estimated with an assumed mass loss rate.  Including the adiabatic cooling in the pressure coordinate system according to \citet{Bildsten}, the entropy equation takes the following form,

\begin{equation}
  \label{eq:9}
  \sum H - \sum C = \frac{C_p \dot{M} T \rho g_p}{4\pi R_p^2 P}\left(\nabla_{\rm ad}-\frac{\dd \ln T}{\dd \ln P}\right),
\end{equation}
where $\sum H$ and $\sum C$ stand for the sum of heating and cooling rates respectively as shown in the right and left hand side of Equation~\ref{eq:heat}, $\nabla_{\rm ad}=(\dd \ln T/\dd \ln P)_S$  is the adiabatic temperature gradient at constant entropy, $\dot{M}$ is the  mass loss rate, and $C_p$ is the specific heat per unit mass at constant pressure.
In the fiducial model, the temperature gradient is about $\dd T/\dd r=5\times 10^{-7}~\rm K~cm^{-1}$ at the radius that 13.6 eV photon becomes optically thick, where most LyC photons get absorbed.  Applying the mass-loss rate $\dot{M}=4\times 10^9 \rm~g~s^{-1}$ suggested by \citet{Salz} using a hydrodynamic escaping atmosphere model, the adiabatic cooling rate is $1.1\times 10^{-8} \ergccs$ and the first term in the Equation~\ref{eq:9} is the dominant source.
Compared to the heating and cooling rates shown in Figure~\ref{fig:cooling}, the adiabatic cooling is more than two orders of magnitude smaller in the region where \Ha\ mostly absorbed, and may only become important in the region above $P\sim 3\times 10^{-4}\rm~\mu bar$.  The mass-loss rate may be model dependent.  An upper bound for the mass loss rate can be found using the energy-limited escape rate \citep{Murray-Clay}, which assumes all LyC flux converts to unbinding the atmosphere.
The LyC flux $F_{\rm LyC}=2.6 \times 10^4 \ergcs$ corresponds to the energy-limited mass-loss rate $\dot{M}=3\times 10^{11} \rm~g~s^{-1}$.  In this case, the adiabatic cooling rate is $8\times 10^{-7} \ergccs$ and still has a less than 15\% effect in the region mainly concerned.

\subsection{Comparing with Other Hot Jupiter Upper Atmosphere Models}
\paragraph{\citet{Christie}}

The present study agrees with the conclusion in \citet{Christie} that the $2p$ occupation is set by radiative excitation and de-excitation, and as an improvement, we include a \Lya\ radiation transfer inside the atmosphere instead of applying a constant solar \Lya\ intensity.  Because of the $\sim 30$ times stronger stellar \Lya\ intensity of HD 189733 comparing to the Sun and considering the \Lya\ photons generated inside the atmosphere due to collisional excitation and recombination cascades, the \Lya\ mean intensity should be $\sim 100$ times larger for the majority of the atomic layer.  In addition, because \citet{Christie} underestimates the $n_{2p}$ by a factor of 20 due to a math error, the $n_{2p}$ should be significantly larger in the whole simulation domain, and thus the atmosphere is optically thick to \Ha\ mainly due to the absorption of H($2p$).  The observed \Ha\ absorption width agrees well with an optically thick atmosphere model.

Because of the much larger $n_{2\ell}$, the photoionization of H(2$\ell$) is larger compared to photoionization of the ground state in the atomic layer.  Hence the $n_e$ and $n_p$ in this work is $\sim 10$ times larger.

Considering the proton collisional $\ell$-mixing process with rate $\sim 10$ times larger than electron collisional process, as well as the large H(2p) population, the creation of 2s hydrogen is dominated by $\ell$-mixing rather than collisional excitation considered in \citet{Christie}.  As a result, $2s$ and $2p$ reach collisional equilibrium.

In addition, it is shown that the metal lines are crucial in cooling the atmosphere.  Assuming solar abundance, lines of Mg and Na can cool the atomic layer by $\simeq 2000-3000$ K.  

\paragraph{\citet{Menager}}

\citet{Menager} investigate the \Lya\ emission and reflection by the atmosphere of \HD.  The temperature and electron, hydrogen, and helium number density profiles of \HD\ from the \citet{Koskinen2011} unpublished model were applied.  The temperature, $n_e$, $n_p$, and $n_{\rm H}$ are in broad agreement with the profiles presented in this paper in the corresponding pressure range.  According to a similar model of HD 209458b presented in \citet{KoskinenI}, it should be a one-dimensional hydrodynamic model of the upper atmosphere considering hydrogen and helium constructed on top of a full photochemical model of the lower atmosphere.  They chose the average solar flux as their stellar spectrum, which is $\sim10$ times smaller than the synthetic spectrum from MUSCLES.  Different from the photoelectron heating efficiency $\eta(E)$ calculated at the fixed ionization fraction $x_e=0.1$ throughout the model, or a constant $\eta$ applied in the \citet{KoskinenI}, a $\eta$ based on the local $x_e$ is used in this work.  

The temperature in the \HD\ model of \citet{Koskinen2011} reachs a peak of about 13000 K at a pressure of $3\times 10^{-4}~\rm \mu bar$.  The adiabatic cooling lowers the temperature at higher altitude.  Their temperature at pressure range $10^{-3}$ to 1 $\mu$bar is higher by about 3000 K.  Two possible reasons of this difference are \citet{KoskinenI} do not consider metal lines cooling, which are the dominant cooling mechanisms in our model, and conduction is not included in this work, which is a net heating in this pressure range according to their result.  Their temperature decreases much faster with pressure above 1 $\mu$bar comparing to Figure~\ref{fig:T}.  The lack of molecular cooling in this work is the possible reason.  Their $n_e$ decreases slowly with pressure in the atomic layer and is $\sim 10$ times smaller at 10 $\mu$bar compared to this work.  This difference is the result of missing the photoionization from H($2s$), which is the dominant ionization mechanism in this region. 

In the \Lya\ radiation transfer simulation of \citet{Menager}, the \Lya\ photons emitted by the star and by the planetary atmosphere are considered.  However, when considering the \Lya\ photon emitted by the planetary atmosphere, the \Lya\ photon created through recombination cascades is not included.   In order to calculate the \Lya\ thermal excitation in the atmosphere, the number densities of $2s$ and $2p$ states hydrogen are modeled with a level population study.  In the $2p$ state level equation, \Lya\ excitation, which completely dominates the $2p$ state, is not included.  In addition, because the $p$ collisional $\ell$-mixing process is missing in their model, the 2$\ell$ state number densities result shown are affected.  

\citet{Menager} claimed that the thermal emission of \HD\ contributes to 6\% of the total intensity of the \Lya\ line.
In the fiducial model of this work, this ratio is $9.6\times 10^3\ergcs / (2.0\times 10^4 \ergcs + 9.6\times 10^3\ergcs) = 32\%$.  This ratio strongly depends on the metallicity as well as the stellar LyC to \Lya\ flux ratio. 

\section{CONCLUSION}

A detailed one-dimensional hydrostatic atmosphere model is constructed over the region dominated by atomic hydrogen and comparison of model transmission spectra to the data has been made.
An atomic hydrogen level population calculation and a Monte-Carlo \Lya\ radiation transfer are done to model the abundance of 2$\ell$ state hydrogen.  The model transmission spectra of \Ha, \Hb, and Na are in broad agreement with the HD 189733b data for both the line center absorption depth and the line width, although the comparison is complicated by the observed variability.  

The \Lya\ radiation transfer shows that the \Lya\ has a very broad line width with a flat top due to the resonant scattering process.
The line profile weighted mean intensity $\bar{J}_{Ly\alpha}$ is large and approximately constant down to the $P=0.1~\rm \mu bar$ level of the atmosphere.
\Lya\ photons created inside the atmosphere and incident from the star are both important.  The \Lya\ source function extends deep into the atmosphere due to ionization from progressively higher energy stellar LyC photons.  The stellar \Lya\ photon can penetrate into very large line center optical depth because the stellar \Lya\ intensity is much boarder than the Doppler width inside the atmosphere.  The stellar \Lya\ photons incident through the surface and the photons generated above $10^{-2}~\rm \mu bar$ can mostly escape through the top boundary.  In contrast, the photons emitted below $0.1~\rm \mu bar$ are mostly absorbed during the resonant scattering processes due to the high optical depth.  For $P\gtrsim 0.1~\rm \mu bar$, $\bar{J}_{Ly\alpha}\propto P^{-1}$.

The $n_{2p}$ is determined by the radiative rates between $1s$ and $2p$ throughout the simulation domain because of the large \Lya\ intensity.  The $2s$ and $2p$ states reach collisional equilibrium by the large $p$ collisional $\ell$-mixing rate, which was overlooked in this context.  The combination of the decreasing \Lya\ excitation rates and the increasing hydrogen density gives rise to a nearly flat $n_{2\ell}$ over two decades in pressure.  This layer is optically thick to \Ha, and the temperature is in the range $T \simeq 3000-8500\, \rm K$.  Both \Ha\ and NaD are optically thick up to the level $P\sim 10^{-2}~\rm \mu bar$, which corresponds to the atomic layer of the atmosphere.  Assuming solar abundance, radiative cooling due to metal species dominates over the entire model, with Mg and Na being the two most important species.  The model shows that \ion{Mg}{2} may have a very large transit depth assuming solar abundance, which might be a good target to constrain the atmospheric properties.

Additional models computed for a range of the stellar LyC flux find transit depth of \Ha\ changes with LyC level, suggesting that the variability in \Ha\ transit depth may be due to variability in the stellar LyC.  In contrast, the Na absorption profile is insensitive to the LyC level.  Since metal lines provide the dominant cooling of this part of the atmosphere, the atmosphere structure is sensitive to the density of species such as Mg and Na, which may themselves be constrained by observations.  Lastly, since the \Ha\ and Na D lines have comparable absorption depths for the same spectral resolution, we argue that the center of the Na D lines are also formed in the atomic layer where the \Ha\ line is formed.  

The present model is in agreement with the observed Na D transmission spectrum by \citet{Huitson} and \citet{Wyttenbach}, although the inferred atmospheric temperature is significantly larger than that found assuming an isothermal profile and molecular composition.  
It is shown that the temperature achieved by fitting each wavelength interval in the observed transmission spectrum with an isothermal atmosphere model may not accurately retrieve the original temperature profile, if the temperature increases rapidly with the altitude.

\section{ACKNOWLEDGMENTS}

We are grateful to Wilson Cauley and Aur\'elien Wyttenbach for kindly providing the data.  We thank Wilson Cauley, Seth Redfield, and Adam Jansen for useful discussions regarding \Ha\ observations.  We also thank the helpful conversations with Ira Wasserman, Craig Sarazin, Shane Davis, Roger Chevalier, and Remy Indebetouw regarding the model construction, and the help from Katherine Holcomb in improving the code efficiency.  The simulations in this work were carried out on the Rivanna computer cluster at the University of Virginia.  We also thank the referee for providing constructive comments and suggestions, especially in suggesting use of the MUSCLES spectrum.  This research was supported by NASA grants NNX14AE16G, NNX10AH29G, and NNX15AE05G.

\appendix

\begin{table*}
\begin{center}
  \caption{List of minor metal cooling transitions}
  \label{tab:cooling_minor}
\begin{tabular}{*6c}
\hline  \hline
Transition & $\Delta E$ & $E_u$ & $A_{ul}$\footnote{\citet{NIST}} & $\Lambda$  & Source of collision rate\\
 & (eV) & (eV) & ($\rm s^{-1}$) & ($\ergcps$) & or cooling rate\\
\hline
O\UI\ 6300+6363 & 1.957 & 1.967 & $7.45\times 10^{-3}$ & $1.3\times 10^{-14}/n_e e^{-2.28/T_4}$\footnote{$T_4=T/10^4{\rm~K}$} & Equation~\ref{eq:11} \\
C\UI\ 9850+9824 & 1.260 & 1.264 & $1.45\times 10^{-4}$ & $3.3\times 10^{-16}/n_e e^{-1.47/T_4}$ & Equation~\ref{eq:11} \\
C\UI\ 8727 & 1.420  & 2.684 & 0.599 & $3.6\times 10^{-21}e^{-3.12/T_4}/(0.599 + 2.39\times 10^{-8} n_e)$&  \citet{Pequignot1976} \\
Si\UII\ 2350+2334 & 5.294 & 5.310 & $1.02 \times 10^3$ & $8.48\times 10^{-12} C_{lu}^{(e)}$ & CHIANTI\\
Si\UII\ 2344 & 5.287 & 5.323 & $1.31 \times 10^3$ & $8.47\times 10^{-12} C_{lu}^{(e)}$ & CHIANTI\\
Si\UII\ 2335 & 5.309 & 5.345 & $2.44 \times 10^3$ & $8.51\times 10^{-12} C_{lu}^{(e)}$ & CHIANTI\\
Si\UII\ 1817($^2D_{3/2}$)+1808 & 6.839 & 6.857 & $2.86 \times 10^6$ & $1.10\times 10^{-11} C_{lu}^{(e)}$ & CHIANTI\\
Si\UII\ 1817($^2D_{5/2}$) & 6.823 & 6.859 & $2.65 \times 10^6$ & $1.10\times 10^{-11} C_{lu}^{(e)}$ & CHIANTI\\
Si\UI\ 16454+16068 & 0.762 & 0.781 & $2.72 \times 10^{-3}$ & $1.84\times 10^{-15}/n_e e^{-0.906/T_4}$ &   Equation~\ref{eq:11} \\
Si\UI\ 10991 & 1.128 & 1.909 & 1.00 & $2.01\times 10^{-13}/n_e e^{-2.22/T_4}$ & Equation~\ref{eq:11}  \\
Si\UI\ 6527 & 1.899 & 1.909 & $2.74 \times 10^{-2}$ & $9.26\times 10^{-15}/n_e e^{-2.22/T_4}$ &  Equation~\ref{eq:11} \\
Si\UI\ 3020+3007 & 4.108 & 4.132 & $4.4 \times 10^3$ & $9.3\times 10^{-24}T^{0.18} e^{-4.79/T_4}$ & Van Regemorter formula\\
Si\UI\ 2988 & 4.149 & 4.930 & $2.66 \times 10^6$ & $4.2\times 10^{-22}T^{0.18} e^{-5.72/T_4}$  & Van Regemorter formula\\
Si\UI\ 2882 & 4.301 & 5.082 & $2.17 \times 10^8$ & $3.1\times 10^{-20}T^{0.18} e^{-5.90/T_4}$ & Van Regemorter formula\\
Si\UI\ 2529+2519+2514 & 4.917 & 4.930 & $2.19 \times 10^8$ & $5.5\times 10^{-20}T^{0.18} e^{-5.72/T_4}$ & Van Regemorter formula\\
Si\UI\ 2524 & 4.911 & 4.920 & $2.22 \times 10^8$ & $1.2\times 10^{-20}T^{0.18} e^{-5.71/T_4}$ & Van Regemorter formula\\
Si\UI\ 2516+2507 & 4.935 & 4.954 & $2.23 \times 10^8$ & $4.3\times 10^{-20}T^{0.18} e^{-5.75/T_4}$ & Van Regemorter formula\\
S\UI\ 25.25$~\mu m$ & 0.0491 & 0.0491 & $1.40 \times 10^{-3}$ & $3.67\times 10^{-17}/n_e e^{-0.0570/T_4}$ & Equation~\ref{eq:11} \\
S\UI\ 11306+10821 & 1.121 & 1.145 & $2.75 \times 10^{-2}$ & $2.78\times 10^{-14}/n_e e^{-1.33/T_4}$ & Equation~\ref{eq:11} \\
S\UI\ 7725 & 1.605 & 2.750 & 1.38 & $3.94\times 10^{-13}/n_e e^{-3.19/T_4}$ & Equation~\ref{eq:11} \\
S\UII\ 10336 & 1.199 & 3.041 & 0.142 & $5.46\times 10^{-14}/n_e e^{-3.53/T_4}$ & Equation~\ref{eq:11} \\
S\UII\ 10320+10287 & 1.203 & 3.046 & 0.272 & $2.09\times 10^{-13}/n_e e^{-3.53/T_4}$ & Equation~\ref{eq:11} \\
S\UII\ 6731 & 1.842 & 1.842 & $6.84 \times 10^{-4}$ & $8.07\times 10^{-16}/n_e e^{-2.14/T_4}$ & Equation~\ref{eq:11} \\
S\UII\ 6716 & 1.845 & 1.845 & $2.02 \times 10^{-4}$ & $3.59\times 10^{-16}/n_e e^{-2.14/T_4}$ & Equation~\ref{eq:11} \\
S\UII\ 4076 & 3.041 & 3.041 & $7.72 \times 10^{-2}$ & $8.27\times 10^{-14}/n_e e^{-3.53/T_4}$ & Equation~\ref{eq:11} \\
S\UII\ 4069 & 3.046 & 3.046 & 0.192 & $3.75\times 10^{-13}/n_e e^{-3.53/T_4}$ & Equation~\ref{eq:11} \\
\hline
\end{tabular}
\end{center}
\end{table*}

Table~\ref{tab:cooling_minor} contains the transitions from O, C, S, and Si lines, which are included in the model as cooling processes but only have a minor effect on the temperature.

\end{CJK*}
\end{document}